\documentclass[reqno,12pt]{amsart}

\usepackage{color}
\usepackage[latin1]{inputenc}
\usepackage[T1]{fontenc}
\usepackage[english]{babel}
\usepackage{amsfonts}
\usepackage{amssymb}
\usepackage{graphicx}
\usepackage{dsfont}
\usepackage{epsfig}
\usepackage{ae}
\usepackage{hyperref}
\usepackage{array}
\usepackage{amsmath}
\usepackage[comma, authoryear]{natbib}
\usepackage{epstopdf}
\usepackage{subfigure}
\usepackage[table,xcdraw]{xcolor}
\usepackage{multirow}
\usepackage{enumitem}
\usepackage{algorithm}
\usepackage{algpseudocode}

\usepackage{amsthm}
\usepackage{appendix}

\newfloat{subroutinea}{htbp}{loa}
\floatname{subroutinea}{Subroutine}
\makeatletter\newcommand\l@subroutinea{\@dottedtocline{1}{1.5em}{2.3em}}\makeatother

\newfloat{subroutineb}{htbp}{loa}
\floatname{subroutineb}{Subroutine}
\makeatletter\newcommand\l@subroutineb{\@dottedtocline{1}{1.5em}{2.3em}}\makeatother

\newcommand{\E}{E}
\newcommand{\V}{\mathds{V}}

\newcommand{\R}{\mathbb{R}}

\newcommand{\F}{\mathcal{F}}

\renewcommand{\(}{\left(}
\renewcommand{\)}{\right)}
\renewcommand{\[}{\left[}
\renewcommand{\]}{\right]}

\newcommand{\1}{\mathds{1}}

\newcommand{\bpm}{\begin{pmatrix}}
\newcommand{\epm}{\end{pmatrix}}

\renewcommand{\hat}{\widehat}
\renewcommand{\proof}{{\noindent \it Proof. }}

\newtheorem{theo}{Theorem}[section]
\newtheorem{pr}{Proposition}[section]
\newtheorem{lem}{Lemma}[section]
\newtheorem{co}{Corollary}[section]
\newtheorem{defi}{Definition}[section]
\newtheorem{re}{Remark}[section]

\newcommand{\be}{\begin{eqnarray}}
\newcommand{\ee}{\end{eqnarray}}
\newcommand{\by}{\begin{eqnarray*}}
\newcommand{\ey}{\end{eqnarray*}}
\newcommand{\bt}{\begin{theo}}
\newcommand{\et}{\end{theo}}
\newcommand{\bl}{\begin{lem}}
\newcommand{\el}{\end{lem}}
\newtheorem*{lem1}{Lemma A.1}

\newtheorem*{lem2}{Lemma A.2}

\newtheorem*{lem3}{Lemma A.3}

\newcommand{\bc}{\begin{co}}
\newcommand{\ec}{\end{co}}
\newcommand{\eex}{\end{exa}\vspace{-3mm}}
\newcommand{\br}{\begin{re}}
\newcommand{\er}{\end{re}\vspace{-3mm}}
\renewcommand{\geq}{\geqslant}
\renewcommand{\leq}{\leqslant}
\renewcommand{\ge}{\geqslant}
\renewcommand{\le}{\leqslant}
\renewcommand{\tilde}{\widetilde}

\algnewcommand\algorithmicto{\textbf{to}}
\algrenewtext{For}[3]%
{\algorithmicfor\ #1 = #2 \algorithmicto\ #3 \algorithmicdo}

\makeatletter
\def\BState{\State\hskip-\ALG@thistlm}
\makeatother

\begin{document}
	
\title{VIX-linked fees for GMWBs via Explicit Solution Simulation Methods}

\author[M. Kouritzin]{Michael A.  Kouritzin}
\address{Department of Mathematical and Statistical Sciences\\
	University of Alberta  \\
	Edmonton (Alberta)\\
	Canada T6G 2G1}
\email{michaelk@ualberta.ca
	\newline\indent {\it URL:} http://www.math.ualberta.ca/Kouritzin\_M.html}

\author[A. MacKay]{Anne MacKay}
\address{Department of Mathematics\\
	Universit\'{e} du Qu\'{e}bec \`{a} Montr\'{e}al \\
	Montreal (Quebec)\\
	Canada H3C 3P8}
\email{mackay.anne@uqam.ca}

\date{\today}

\thanks{Partial funding in support of this work
	was provided by an NSERC discovery grant.}

\renewcommand{\subjclassname}{\textup{2010} Mathematics Subject Classification}

\begin{abstract}

	In a market with stochastic volatility and jumps, we consider a VIX-linked fee structure (see \cite{cui2016VIX}) for variable annuity contracts with guaranteed minimum withdrawal benefits (GMWB). Our goal is to assess the effectiveness of the VIX-linked fee structure in decreasing the sensitivity of the insurer's liability to volatility risk. Since the GMWB payoff is highly path-dependent, it is particularly sensitive to volatility risk, and can also be challenging to price, especially in the presence of the VIX-linked fee. In this paper, following \cite{kouritzin2016path}, we present an explicit weak solution for the value of the VA account and use it in Monte Carlo simulations to value the GMWB guarantee. Numerical examples are provided to analyze the impact of the VIX-linked fee on the sensitivity of the liability to changes in market volatility.
	
\end{abstract}

\maketitle

\section{Introduction}

Insurers and other financial institutions offering equity-linked insurance products are exposed to different financial, demographic and behavioral risks. In particular, the guarantees embedded in variable annuities (VA) are linked to the performance of financial markets, and they depend on the survival of the policyholder. Withdrawal, lapse and surrender behavior of the policyholder also have a significant impact on the revenue, the liabilities and the risk profile of the insurer. In order to reduce the exposure of life insurers to market volatility risk, the Chicago Board of Options Exchange (CBOE) has suggested linking the VA insurance fee rate to the VIX index \citep{CBOE,CBOE2}, which is considered a proxy for short-term market volatility.

Volatility risk has a significant impact on insurers that hedge VA liabilities using market-traded derivatives. VA guarantees typically have a long horizon, which can extend over twenty years, while market-traded options are usually not liquid past the one-year mark. A replicating portfolio for VA guarantees is therefore composed of shorter-term instruments that need to be refreshed frequently. High market volatility increases option prices, and therefore impacts the cost and efficiency of the insurer's hedging strategy.

Risk management of a variable annuity can also be complicated by the fact that the embedded guarantees are financed via an insurance fee, which is typically quoted as a fixed percentage of the underlying VA account. These fee withdrawals reduce the actual return on the account, much like a management fee in mutual funds. A fixed fee structure creates a discrepancy between the fee income and the liability associated with the embedded guarantee. Indeed, the fee income decreases when the VA account value drops, but such a drop increases the value of the guarantee. Since equity returns and market volatility tend to be negatively correlated (the so-called \textit{leverage effect}), linking the fee rate to the VIX index helps realign the fee income and the liability.

A VIX-linked fee structure is analyzed in \cite{cui2016VIX}, who jointly describe the dynamics of the variable annuity account value and the VIX index. Using techniques based on characteristic function inversion, the authors calculate the net liability associated with a guaranteed minimum accumulation benefit (GMAB) when the market follows the Heston model (\cite{H93}). They show that the VIX-linked fee is in fact effective in reducing the sensitivity of the net liability to temporary variations in market volatility. In other words, when a VIX-linked fee rate is used, the impact on the net liability of an increase in market volatility is not as significant as when the GMAB is financed via a fixed-fee rate. A similar VIX-linked fee is also studied by \cite{bernardvariable}, who link the VIX index to S\&P 500 prices using a Gaussian copula.

There has been a recent interest in the application of market-dependent fee structures, which include VIX-linked fee, to finance VA guarantees. For example, \cite{BHM14} propose a fee paid only when the account value is below a certain threshold. This fee structure is analyzed in terms of surrender incentive behavior in \cite{mackay2017risk}, of fee deduction periods in \cite{zhou2015time} and is extended to more general market models in \cite{D14}. 

Hitherto, in the academic literature, market-dependent fees have only been applied to accumulation benefits. In this paper, we focus on guaranteed minimum withdrawal benefits (GMWB). In exchange for an initial premium, a GMWB guarantees the policyholder a minimal income for a given number of years, regardless of market performance. The payoff resulting from this type of guarantee has path-dependent features, and analytic forms for the associated liability are not always possible, especially in market models that include stochastic volatility and jumps.

GMWBs are often studied assuming either deterministic withdrawals, or ``optimal'' (i.e. worst-case scenario for the insurer) withdrawals. For examples of the first assumption, see \cite{milevsky2006financial}, \cite{bauer2008universal} and \cite{feng2017risk}. \cite{chen2008effect}, \cite{dai2008guaranteed} and \cite{luo2015valuation} are examples of articles considering optimal withdrawals. In all cases, the insurance fee rate is assumed to be fix. The present paper is closer to the first category, but the framework we present can accommodate both deterministic and adapted withdrawals. This allows us to focus on volatility risk. Analysis of the impact of the VIX-linked fee on the optimal policyholder behavior is left for future work.

In this paper, we consider a market with stochastic volatility and jumps, for which we assess the impact of the VIX-linked fee structure on the net liability and economic capital of a GMWB using Monte Carlo simulations. More precisely, we show that a VIX-linked fee structure decreases the sensitivity of the net liability to variations in market volatility. We use simulation method based on an explicit (weak) solution of the stochastic differential equation (SDE) describing the value of the VA account. This method was first introduced in \cite{kouritzin2016path} and is based on the results of \cite{Kouritzin/Remillard:2015}. Since it stems from an exact solution rather than on the transition density of the spot volatility, our method is very flexible and could be extended to applications beyond the one presented here. It also avoids the issues resulting from Euler or Milstein discretization, such as a slow convergence rate and the possibility of simulations producing negative volatility.

Explicit solutions to non{-}linear stochastic differential equations are rare.  However, when they arise and have a simple form, they can be most useful.  There are two types of explicit solutions: strong and weak.  Strong solutions are adapted to the driving processes and their explicit representation (if any) would be in terms of  these driving processes only.  Explicit weak solutions involve {\em extra randomness}, which does not show in the stochastic differential equations, but are equally useful as strong solutions if this extra randomness does not consume significant computational time to produce or if it can be produced offline. \citet{kouritzin2016path} showed the Heston model has explicit weak solutions, which are obtained as projections of strong solutions to a higher dimensional stochastic differential equations of the form studied by \citet{Kouritzin/Remillard:2015}.  Such a Heston weak solution could be used to recover known marginal (and unknown joint) distributions at various times but more importantly gives an approximation{-}free formula to base simulation upon.  Herein, we will show that the generalized Heston model used to describe the value of the VA account also yields weak solutions.  Further, it will be explained in the appendix that the weak{-}solution {\em extra randomness} can be simulated offline.

The main contribution of this paper is twofold. First, we extend the work of \cite{cui2016VIX} by applying the VIX-linked fee to a GMWB, and by adding jumps to the index value process. In this context, we show that the net liability associated with the GMWB rider is less sensitive to changes in market volatility when the fee rate is a function of the VIX index. Hence, the conclusions of \cite{cui2016VIX} regarding GMABs in a stochastic volatility market can be extended to GMWBs in a more general market model. 

In this paper, we also obtain an explicit weak solution for the value of the variable annuity account in an extended Heston market model with jumps, and use it to propose a new simulation method. We provide the details of the simulation algorithm, and we suggest modifications based on branching particles, which improves the performance of the algorithm. While the solution to the extended Heston-like SDE is new, the simulation algorithms are based on the ideas presented in \cite{kouritzin2016path} and \cite{kouritzin2017branching}.

The rest of this article is as follows. In Section \ref{sec:2_financial_model}, we fix the financial setting of our work, and we introduce GMWB contract and the valuation framework. Section \ref{sec:3_simulation} contains the main theoretical results and the resulting simulation method. This method is used to present numerical examples in Section \ref{sec:4_numerical_results}, and Section \ref{sec:5_conclusion} lists the conclusions. More involved proofs, as well as detailed simulation algorithms are provided in appendices.

\section{Financial model and variable annuity contract}\label{sec:2_financial_model}

\subsection{Variable annuity account}

We consider the simplified situation where the variable annuity (VA) account is invested in a single fund, which tracks an equity index. In this paper, as in \cite{CBOE,CBOE2} and \cite{cui2016VIX}, we assume that this index is the S\&P 500. 

Let $T \in (0,\infty)$ be the final time of the model. It could, for example, represent the last time at which the cash flows coming from the VA contract can be paid; no financial modeling is required beyond that point. Throughout this paper, we use the following notation:
\begin{itemize}
	\item $\{S_t\}_{0 \leq t \leq T}$ is the value of the equity index;
	\item $\{F_t\}_{0 \leq t \leq T}$ is the value of the variable annuity account.
\end{itemize}

\subsubsection{Equity index}

We consider a probability space $(\Omega,\F,\mathbb{F},P)$, where $P$ denotes the real-world (objective) measure. We model the index price process using a model that incorporates stochastic volatility and jumps (see for example \cite{bates2000post}, \cite{andersen2002empirical}, and \cite{pan2002jump}), and we refer to it as the stochastic volatility with jumps (SVJ) model. We therefore assume that $\{S_t\}_{0 \leq t \leq T}$ has the following $P$-measure dynamics:
\begin{align}
d\bpm S_t \\ V_t \epm
= \bpm (r + \eta_S V_t - \lambda^* \delta^*) S_t \\ \nu-\varrho^* V_t \epm
dt +
\bpm \sqrt{1-\rho^2} S_t V^{\frac{1}{2}}_t & \rho S_t V^{\frac{1}{2}}_t & S_{t^-} \\
0 & \kappa V^{\frac{1}{2}}_t & 0\epm
\bpm dB^*_t\\ d\beta^*_t \\ dX^*_t\epm
\label{hestonP}
\end{align}
for $0 \leq t \leq T$, with $S_0=s>0$, $V_0 = v_0 > 0$,  $\nu>0$, $\varrho^*>0$, $\kappa>0$ and $-1 \leq \rho \leq 1$. We let $r \geq 0$ and $\eta_S \geq 0$ denote the risk-free rate and the equity risk premium, respectively. $B^*_t$ and $\beta^*_t$ are $P$-Brownian motions, and $X^*_t = \sum_{i=1}^{N^*_t} J^*_i$ is a compound Poisson process, with $\{N^*_t, t>0\}$ a Poisson process of intensity $\lambda^*$. For ${i = 1,\ldots,\infty}$, we let ${J^*_i = e^{Y^*_i}-1}$, and where each $Y^*_i$ follows a Normal distribution with mean ${(\log(1+\delta^*)-\frac{\chi^2}{2})}$ and variance $\chi^2$. We assume that  $B^*_t$, $\beta^*_t$, $N^*_t$ and $Y^*_i$, $i=1,\ldots,\infty$ are mutually independent.

From the definition of $J^*_i$, it follows that ${E[J^*_i] = \delta^*}$, and that the magnitude of jumps $J^*_i$ is greater than $-1$. This ensures that the price process does not reach $0$ when a jump occurs.

\begin{re}
	The model above is usually written in terms of the correlated $P$-Brownian motions $\tilde{B}^*_t$ and $\tilde{\beta}^*_t$, with $\langle \tilde{B}^*, \tilde{\beta}^* \rangle_t = \rho t$:
	\begin{align}
	d
	\bpm S_t \\ V_t \epm
	= \bpm (r + \eta_S V_t -\lambda^* \delta^*) S_t \\ \nu-\varrho^* V_t \epm
	dt +
	\bpm S_t V^{\frac{1}{2}}_t & 0 & S_{t^-}\\
	0 & \kappa V^{\frac{1}{2}}_t & 0 \epm
	\bpm d\tilde{B}^*_t\\ d\tilde{\beta}^*_t \\ dX^*_t\epm.
	\label{hestonPalt}
	\end{align}
	Representations \eqref{hestonP} and \eqref{hestonPalt} are equivalent, and our use of \eqref{hestonP} facilitates the explicit representation of the dependence between randomness of the index price and the randomness of the volatility process.
\end{re}

The introduction of stochastic volatility and jumps make the market described above incomplete, and thus there exist infinitely many equivalent local martingale measures. In this work, we follow \cite{duan2010jump} for the specification of the market prices of risk. In particular, we denote the market price of volatility by $\eta_V$, and we use composite parameters $\phi$ and $\phi^*$ to define the market price of jump risk. We define
\begin{align}
\phi &= \lambda\(\delta - \log(1+\delta) + \frac{\chi^2}{2}\),\label{eq:phi}\\
\phi^* &= \lambda^*\(\delta^* - \log(1+\delta^*) + \frac{\chi^2}{2}\) \nonumber
\end{align}
and let $\eta_J = \phi - \phi^*$ be the jump risk premium. Therefore, the jump process can have different intensities and mean size of jumps under each measure (see Section 3 of \cite{duan2010jump} for more details). Under these assumptions, the dynamics of the equity index and the volatility process keep a similar form under the risk-neutral measure, but have different parameters. Thus, under the pricing risk-neutral measure $Q$, we have

\begin{align}
d
\bpm S_t \\ V_t \epm
= \bpm (r - \delta \lambda) S_t \\ \nu-\varrho V_t \epm
dt +
\bpm \sqrt{1-\rho^2} S_t V^{\frac{1}{2}}_t & \rho S_t V^{\frac{1}{2}}_t & S_{t^-}\\
0 & \kappa V^{\frac{1}{2}}_t & 0 \epm
\bpm dB_t\\ d\beta_t \\ dX_t\epm,
\label{heston}
\end{align}
where $r$ is the risk-free interest rate and $\varrho=\varrho^*+\eta_V$. $B_t$ and $\beta_t$, given by
\begin{align*}
B_t &= B^*_t + \frac {\kappa \eta_S + \rho \eta_V}{\kappa \sqrt{(1-\rho^2)}} \int_0^t V_s^{\frac12} ~ ds,\\
\beta_t &= \beta^*_t + \frac{\eta_V}{\kappa} \int_0^t V^{\frac12}_s ~ ds,
\end{align*}
are $Q$-Brownian motions. $X_t = \sum_{i=1}^{N_t} J_i$ with $\{N_t, t > 0\}$ is a Poisson process of intensity $\lambda$ and for $i=1,\ldots$, $J_i=e^{Y_i}-1$, with each $Y_i$ normally distributed with mean ${(\log(1+\delta)-\frac{\chi^2}{2})}$ and variance $\chi^2$. We also have that $B_t$, $\beta_t$, $N_t$ and $\{Y_i\}_{i=1}^{\infty}$ are mutually independent. 

\begin{re}
	The SVJ model presented above includes the \cite{H93} model, which was used in a similar context in \cite{cui2016VIX}, as a special case. To obtain results in the Heston model, it suffices to set the parameters linked to the compound Poisson process to $0$.
\end{re}

\subsubsection{The variable annuity fee schedule}

The fee rate paid out of the VA account is composed of two parts. The first part is the \emph{investment management fee}. It is paid to the third party who manages the underlying investment fund. This part of the fee is not returned to the insurer who offers the financial guarantees. 

The second part of the fee is called the \emph{rider charge}, or the \emph{rider fee}. It goes to the insurer and covers the cost of the financial guarantee, or rider.

We assume that the investment management fee is fixed, and we denote it by $q$. Only the rider fee, denoted by $c_t$, will be linked to the VIX index. We denote the total fee rate payable at time $t$ by $\gamma_t$, and define it as the sum of the two types of fees:
\begin{equation}
\gamma_t = q+c_t.
\label{eq:gammat}
\end{equation}

In this paper, we consider the case where the rider fee is linked to the VIX index. The purpose of the VIX index is to measure the market's expectation of 30-day volatility implied from at-the-money S\&P 500 Index (SPX) option prices. The CBOE calculates the VIX as a weighted average of SPX call and put prices over a wide range of strikes.

As explained in \cite{kokholm2015joint}, under certain assumptions on the price process, the square of the VIX index can be expressed as a function of the risk-neutral expectation of the log contract. These assumptions include continuity of the price process, which is not satisfied in our setting, and therefore, we have that the square of the VIX at time $t$, $\mathrm{VIX}^2_t$ is given by
\begin{equation*}
\mathrm{VIX}_t^2 =\frac{2}{\bar{\tau}}
\E^Q_t\[\log\(\frac{S_{t+\bar{\tau}}}{\bar{S}^{\bar{\tau}}_t}\)\] + \epsilon,
\end{equation*}
with $\bar{\tau}=30/365$, where $\E^Q_t[\cdot]$ is shortened notation for $\E^Q[ \cdot|\mathcal{F}_t]$, $\bar{S}^{\bar{\tau}}_t = S_t e^{r\bar{\tau}}$ denotes the price of a forward contract on the equity index and $\epsilon$ is an error term that is due the presence of jumps in the price process. The error term also accounts for the fact that options are only tradable for a finite number of strikes. As shown in \cite{carr2008variance} (see also  \cite{kokholm2015joint}), in the case of the SVJ model that we consider, this error term is negligible. Going forward, we will ignore it and use:
\begin{equation}
\mathrm{VIX}_t^2 =\frac{2}{\bar{\tau}}
\E^Q_t\[\log\(\frac{S_{t+\bar{\tau}}}{\bar{S}^{\bar{\tau}}_t}\)\].
\label{eq:vix_approx}
\end{equation}
This assumption is common in the VIX derivative literature (see for example \cite{lin2007pricing}, \cite{lian2013pricing} and \cite{kokholm2015joint}). In the SVJ model, the risk-neutral expectation of the log contract can be written as an affine function of the current instantaneous volatility $V_t$:
\begin{equation}
\frac{2}{\bar{\tau}}\E^Q_t\[\log\(\frac{S_{t+\bar{\tau}}}{\bar{S}^{\bar{\tau}}_t}\)\]=A+B V_t,\label{cal1}
\end{equation}
where
\begin{align}
A&=\frac{\nu(\varrho\bar{\tau}-1+e^{-\varrho\bar{\tau}})}{\varrho^2\bar{\tau}} + 2\phi,\nonumber\\ 
B&=\frac{1-e^{-\varrho\bar{\tau}}}{\varrho\bar{\tau}},
\label{eq:AB}
\end{align}
with $\phi$ as defined in \eqref{eq:phi}.
Therefore, from \eqref{eq:vix_approx}, the square of the VIX index at time $t$ can be written as an affine function of $V_t$
\begin{equation}
\mathrm{VIX}_t^2 = A+B V_t.
\label{eq:VIX2}
\end{equation}

The main purpose of linking the fee rate to the VIX index is to re-align the value of the VA guarantee with the fee income, to reduce the sensitivity of the net liability to changes in the volatility. Since VA guarantees are typically more valuable when volatility is high, we seek to achieve positive correlation between the fee income and the VIX. Therefore, we let the fee be a linear function of the square of the VIX:
\begin{align} \label{eq:ct}
c_t&:=\bar{c}+m\mathrm{VIX}^2_t,
\end{align}
where $m>0$ is the ``multiplier'' and $\bar{c}>0$ is the ``base fee rate''. From the definition of the VIX squared, $c_t > 0$ for all $t$ and $V_t$.

It is possible to re-write the total fee rate $\gamma_t$ in terms of the instantaneous volatility. By substituting \eqref{eq:AB}, \eqref{eq:VIX2} and \eqref{eq:ct} into \eqref{eq:gammat}, we obtain
\begin{align} \label{ccb}
\gamma_t&= \alpha_0 + \alpha~V_t,
\end{align}
where 
\begin{align*}
\alpha_0 &= q+\bar{c} + \frac{m \nu}{\varrho}
\left(\frac{\varrho \bar{\tau} -1 + e^{-\varrho \bar{\tau}}}{\varrho \bar{\tau}}\right) + 2\phi,\\
\alpha &= m (1-e^{-\varrho \bar{\tau}})/(\varrho \bar{\tau}),
\end{align*}
with $\phi$ as defined in \eqref{eq:phi}.

\subsubsection{The variable annuity account}\label{ssec:VAaccount}

The variable annuity account we present here is a generalization of the one used in \cite{cui2016VIX}. In addition to the total fee, withdrawn from the account at a rate $\gamma_t$, we consider another cashflow representing withdrawals made by the policyholder. This allows us to price VA policies with a guaranteed minimum withdrawal benefit (GMWB) rider, which lets the policyholder make withdrawals (up to a pre-defined annual amount) until an amount equal to the initial investment has been withdrawn.\footnote{Some GMWB riders can include other features that allow to increase the total amount available for withdrawals. We do not consider those in our paper.} The rate of withdrawal at time $t$ is denoted by $w_t\geq 0$.

Therefore, for $0 \leq t \leq T$ and for $F_t > 0$, the variable annuity account evolves according to
\begin{equation}
\frac{dF_t}{F_t} = \frac{dS_t}{S_t} - \gamma_t dt - \frac{w_t}{F_t} dt,
\label{dFt_pre}
\end{equation}
with $F_0=P$, the amount of the initial premium. Using \eqref{heston} and \eqref{ccb} in \eqref{dFt_pre}, we have, for $F_t>0$,
\begin{align}
d
\bpm F_t \\ V_t \epm
= \bpm \mu F_t -\alpha V_t F_t - w_t \\ \nu-\varrho V_t \epm
dt +
\bpm \sqrt{1-\rho^2} F_t V^{\frac{1}{2}}_t & \rho F_t V^{\frac{1}{2}}_t  & F_{t^-}\\
0 & \kappa V^{\frac{1}{2}}_t & 0\epm
\bpm dB_t\\ d\beta_t \\ dX_t\epm,
\label{eq:dFt}
\end{align}
for $0 \leq t \leq T$, where $\mu = r - \delta\lambda - \alpha_0$. Note that because of the withdrawals $w_t$, it is possible that $F_t$ reaches $0$ (but not at a jump time). In this case, we assume the process is absorbed at $0$ and does not become negative. Thus, for $F_t \leq 0$, $dF_t = 0$.

For risk management purposes, it is important to know the dynamics of the account value under the (objective) $P$-measure. Using \eqref{hestonP} and \eqref{dFt_pre}, we obtain
\begin{align}
d
\bpm F_t \\ V_t \epm
= \bpm \bar{\mu}^* F_t -\alpha^* V_t F_t - w_t \\ \nu-\varrho^* V_t \epm
dt +
\bpm \sqrt{1-\rho^2} F_t V^{\frac{1}{2}}_t & \rho F_t V^{\frac{1}{2}}_t  & F_{t^-}\\
0 & \kappa V^{\frac{1}{2}}_t & 0\epm
\bpm dB^*_t\\ d\beta^*_t \\ dX^*_t\epm,
\label{eq:dFt_P}
\end{align}
with $\bar{\mu}^* = r - \delta^*\lambda^* -\alpha_0$, $\alpha^* = \alpha - \eta_S$ and $B^*_t$, $\beta^*_t$ and $X^*_t$ as in \eqref{hestonP}. 

\begin{re}
	This setting can also be used to model other types of VA accounts. As mentioned earlier, by letting $w_t = 0$ for all $t \in [0,T]$, and by setting the jump parameters to $0$, we recover framework presented in \cite{cui2016VIX}. By letting $w_t$ be negative, we can use it to represent deposits made in the VA account by the investor. This type of deposits are known as \textit{flexible premiums}, and are discussed in \cite{chi2012} and \cite{Bernard2016}. Although we do not focus on them here, the results we present in this paper can easily be used to price guarantees on VAs with flexible premiums.
\end{re}

\subsection{Pricing variable annuity guarantees}\label{ssec:pricingVAs}

In this section, we consider a general variable annuity contract with maturity $T$. Any payout from the financial guarantee will be made before this time. We denote by $\varphi_t(F)$ the payoff of the guarantee at time $t$, $0 \leq t \leq T$, and assume that
\begin{equation*}
{\E\left[\left(\int_0^T \varphi_t(F) dt\right)^2\right]<\infty}.
\end{equation*} 
Note that $\varphi_t(F)$ can depend on the full path of $F$ up to time $t$, not only on its value at $t$, $F_t$. In the next section, we will decribe the specific guarantee in which we are interested, but first, we define the concepts and quantities of interest.

We denote by $\tau_x$ the random variable representing the future lifetime of a policyholder aged $x$ at $t=0$, and we assume that $\tau_x$ is independent of the financial market.

\begin{defi}[Discounted future loss]\label{def:FutureLoss}
	The discounted future loss at time $t$, $0 \leq t \leq T$, is denoted by $\Lambda(t, F, V)$ and is given by
	\begin{equation*}
		\Lambda(t,F,V) = \int_t^T e^{-r(u-t)} \varphi_u(F)\1_{\{\tau_x > u\}} ~ du - \int_t^T e^{-r(u-t)} c_u F_u \1_{\{\tau_x > u\}} ~du.
	\end{equation*}
\end{defi}
The discounted future loss is a random variable representing the difference between the future outflows and inflows to the insurer, for a given guarantee. Both cash flows stop if the policyholder dies before $T$. It is important to note that the fee income to the insurer only takes the rider fee $c_t$ into account, since the investment management part of the fee $q$ is paid to a third party. To simplify the notation, $\Lambda(t,F,V)$ is shortened to $\Lambda_t$ whenever this notation does not cause confusion.

\begin{defi}[Net liability]\label{def:NetLiab}
	The net liability at time $t$, $0 \leq t \leq T$ associated with the financial guarantee described by $\{\varphi_t\}_{0\leq t \leq T}$ is denoted by $\Pi(t,F_t,V_t)$, and is
	given by
	\begin{equation*}
	\Pi(t,F_t,V_t) = E^Q\left[\Lambda_t|\F_t\right].
	\end{equation*}
\end{defi}

The net liability is thus the risk-neutral expectation of the discounted future loss. 

\begin{defi}[Fair fee]\label{def:ffee}
	A vector $(\bar{c}^*,m^*)$ is called a \emph{fair fee structure} (or \emph{fair fee}) if it satisfies
	\begin{equation}
	\Pi(0,F_0,V_0;\bar{c},m) = 0.
	\label{eq:ffee_condition}
	\end{equation}
\end{defi}
In the above equation, we add the parameters $\bar{c}$ and $m$ to the notation of the net liability at $0$, to highlight its dependence on the fee structure \eqref{eq:ct}. From Definition \ref{def:ffee}, a VA contract is fairly priced if the insurer's expected discounted payout to the policyholder is equal to the expected discounted fee income.

For a given contract, the fair fee is generally not unique. More than one vector $(\bar{c},m)$ can satisfy \eqref{eq:ffee_condition}; see Section \ref{sec:4_numerical_results} for an example. To find a fair fee structure, one typically fixes either $\bar{c}$ or $m$ and uses \eqref{eq:ffee_condition} to solve for the other.

\begin{re}
	A common way to express the fair fee condition in the GMWB literature is to set the fee such that the initial value of VA account, $F_0$, is equal to the risk-neutral expected value of the total payout of the contract. \cite{milevsky2006financial} shows that this total payout can be expressed as the sum of a term-certain annuity and a claim on the value of the VA sub-account, and that the resulting fair fee condition is
	\begin{equation*}
	F_0 = \int_0^T e^{-rt} w_t dt + E^Q[e^{-rT}F_T].
	\end{equation*}	
	This particular form of the fair fee condition has since then been widely used to price GMWB contracts, since it simplifies the quantities involved in the calculation. In our setting, the addition of a non-zero third-party management fee $q$ breaks down the equivalence of the fair fee condition obtained by \cite{milevsky2006financial}. Indeed, using the definition of $F_t$ and the optional stopping theorem, it is possible to show that the fair fee condition \eqref{eq:ffee_condition} is equivalent to
	\begin{equation}
	F_0 = \int_0^T e^{-rt} w_t dt + E^Q[e^{-rT}F_T] + E^Q\left[\int_0^\tau qe^{-rt}F_t dt\right].
	\label{eq:ffee_alt}
	\end{equation}
	The additional term on the right-hand side of \eqref{eq:ffee_alt} is strongly path dependent, thus reducing the interest of the alternate form of the fair fee condition. Furthermore, since \eqref{eq:ffee_condition} and \eqref{eq:ffee_alt} are equivalent, they cannot be used together to determine a unique fair fee structure $(\bar{c}^*,m^*)$.
\end{re}

Using \eqref{eq:gammat} and \eqref{ccb}, the risk-neutral present value of future rider fee income can be written as
\begin{align*}
&\E^Q_t\left[\int_t^T e^{-r(u-t)} F_u c_u \1_{\{\tau_x > u\}} du\right] =\nonumber\\
&\qquad(\alpha_0-q)\E^Q_t\left[\int_t^T e^{-r(u-t)} F_u \1_{\{\tau_x > u\}} ~du \right] 
+ \alpha \E^Q_t \left[\int_t^T e^{-r(u-t)} V_u F_u \1_{\{\tau_x > u\}} ~ du\right].
\end{align*}

In the next section, we present the specific guarantee we seek to price.

\subsection{Risk Management of Variable Annuity Guarantees}\label{ssec:riskmeasures}

It is of interest for insurers and regulators to assess the distribution of the discounted future loss $\Lambda_t$. In particular, as explained in \cite{feng2014comparative}, tail risk measurement of the distribution of the initial loss $\Lambda_0$ can be used to determine reserves and risk capital. In this paper, we apply risk measures to the distribution of $\Lambda_0$ to study the effectiveness of the VIX-linked fee in reducing the sensitivity of the risk capital to changes in market volatility.

Two of the most popular tail risk measures in insurance are the value-at-risk (VaR) and the conditional tail expectation (CTE), which we recall in this section. 

\begin{defi}[Value-at-risk]
	Let $\Lambda$ be an $\R$-valued random variable and let $\zeta$ denote a confidence level in $(0,1)$. The value-at-risk at a level $\zeta$, denoted by $VaR_\zeta(\Lambda)$, is defined as
	\begin{equation*}
		\inf \{y \in \R: P(\Lambda\leq y)\geq \zeta\}.
	\end{equation*} 
\end{defi}

When the distribution function of $\Lambda$ is continuous, then the value-at-risk at a level $\zeta$ is simply the $\zeta$-quantile of the distribution.

\begin{defi}[Conditional tail expectation]
	Let $\Lambda$ be an $\R$-valued random variable and let $\zeta$ denote a confidence level in $(0,1)$. If the distribution function of $\Lambda$ is continuous, then the conditional tail expectation, denoted by $CTE_\zeta(\Lambda)$, is defined as
	\begin{equation*}
	CTE_\zeta(\Lambda) = E^P\[\Lambda | \Lambda > VaR_\zeta(\Lambda)\].
	\end{equation*}
\end{defi}

In our market model, the distribution of $\Lambda_t$ is continuous for all $0 \leq t \leq T$. Therefore, it is not limiting to restrict the definition of the CTE to the continuous distribution case.

\begin{re}
	In the context of variable annuities, the risk measures presented here should be calculated under the real-world probability measure, since their purpose is risk-management, rather than pricing.
\end{re}

More details on the use of risk measures in the context of variable annuities can be found in Chapter 9 of \cite{hardy2003investment}.

\subsection{Guaranteed Minimum Withdrawal Benefit}\label{ssec:GMWB}

A guaranteed minimum withdrawal benefit (GMWB) gives the policyholder the right to withdraw from the VA account at a maximum given rate (denoted here by $\bar{w}$) until the full amount of the initial premium $F_0$ is withdrawn, regardless of the current account value. That is, if the account value reaches $0$ before the total of the withdrawals reaches $F_0$, the policyholder can still make withdrawals, and the insurer needs to fund them. 

Here, we consider that $T$ is the maturity of the guarantee, and we let it be defined as the time at which an amount equal to the initial premium has been withdrawn, so that
\begin{equation*}
T=\inf\left\{t>0:\int_0^t w_s ds = F_0\right\}.
\end{equation*}
If we assume that the withdrawals are made at the maximum annual rate $\bar{w}$, then we have  $T=F_0/\bar{w}$. In this paper, we will consider deterministic withdrawal rates that are not necessarily constant.

The guaranteed withdrawals made after the account is depleted represent the payoff of the GMWB rider. If the index tracked by the VA account performs well, it is possible that the account value is still positive at $T$. In this case, the payoff of the GMWB rider is equal to $0$.

We let $\tau_0$ be a stopping time representing the moment at which the VA account value reaches $0$:
\begin{equation}
\tau_0 =: \inf\left\{t>0: F_t = 0\right\}.
\label{eq:tau0}
\end{equation}
Then $\tau=:\tau_0 \wedge T$ is the time at which the GMWB rider is triggered. If the account value is still positive at maturity of the contract, then $\tau=T$. 

Henceforth, in order to focus on the financial risk, we also assume 
\begin{equation*}
P(\tau_x > T) = 1.
\end{equation*}
In other words, the policyholder is assumed to survive at least until maturity of the contract, and for any $t \leq T$, $\1_{\{\tau_x > t\}} = 1$.

The payoff function of the GMWB, $\varphi^W(F)$, can then be written as
\begin{equation}
\varphi^W_t(F) =
\begin{cases}
0, & t < \tau,\\
w_t, & t \geq \tau.
\end{cases}
\label{eq:payoffGMWB}
\end{equation}

We denote the discounted future loss associated with the GMWB rider at time $t$ by $\Lambda^W(t,F,V)$, shortened to $\Lambda^W_t$ when possible. Using \eqref{eq:payoffGMWB} and Definition \ref{def:FutureLoss}, we get
\begin{equation}
	\Lambda^W(t,F,V) = 
	\begin{cases}
	\int_{\tau}^{T}e^{-r(u-t)} w_u ~ du - \int_t^T e^{-r(u-t)} c_u F_u ~ du, & F_t > 0,\\
	\int_{t}^{T}e^{-r(u-t)} w_u ~ du - \int_t^T e^{-r(u-t)} c_u F_u ~ du, & F_t = 0.
	\end{cases}\label{eq:FutureLossW}
\end{equation}

From \eqref{eq:FutureLossW}, it is possible to observe that the randomness in the first term on the right-hand side only comes from $\tau$, the time at which the VA account is depleted, since we assume a deterministic withdrawal rate $w_t$.

\begin{pr}\label{co:liabilityGMWB}
	The net liability associated with the GMWB rider at time $t$, $\Pi^W(t,F,V)$ is given by 
	\begin{align}
	\Pi^W(t,F_t,V_t) &= E_t^Q\[\Lambda^W_t\]\nonumber\\
	&=\begin{cases}
	E^Q_t\[\int_{\tau}^{T}e^{-r(u-t)} w_u ~ du\] - 
		E^Q_t\[\int_t^\tau e^{-r(u-t)}c_u F_u ~ du\], & F_t > 0,\\
	E^Q_t\[\int_t^{T}e^{-r(u-t)} w_u ~ du\], & F_t = 0.
	\end{cases}
	\label{eq:liabilityGMWB}
	\end{align}
\end{pr}

\proof
	The result follows from Definition \ref{def:NetLiab}, by taking the risk-neutral expectation of \eqref{eq:FutureLossW}. Then, since the process $F$ is absorbed at $0$, we have
	\begin{align}
		\E^Q_t\left[\int_t^T e^{-r(u-t)}c_u F_u ~ du\right] 
		&= \E^Q_t\left[\int_t^T e^{-r(u-t)}c_u F_u \1_{\{\tau>u\}} ~ du\right]\nonumber\\
		&= \E^Q_t\left[\int_t^\tau e^{-r(u-t)}c_u F_u ~ du\right].
		\label{eq:net_liability_GMWB}
	\end{align}
	when $F_t>0$, and $\E^Q_t\left[\int_t^T e^{-r(u-t)}c_u F_u ~ du\right] = 0$ when $F_t = 0$.
\qed

The main objective of this paper is to study the net liability $\Pi^W(t,F_t,V_t)$ and the distribution of $\Lambda^W_0$, when the rider fee $c_t$ is linked to the VIX ($m>0$). From \eqref{eq:FutureLossW} and\eqref{eq:liabilityGMWB}, it is clear that these quantities are strongly path-dependent, because of the stopping time $\tau$ and the rider fee $c_t$. Therefore, analytical expressions such as the one presented in \citet{cui2016VIX} are not necessarily available, especially when the withdrawal rate $w_t$ is not constant. We thus estimate risk-measures on the distribution of $\Lambda^W_0$ and calculate $\Pi^W(t,F_t,V_t)$ by simulating the VA account process, using the simulation method developed in \citet{kouritzin2016path}. In the next section, we show how to adapt this method to our setting.

\section{Simulating the variable annuity account}\label{sec:3_simulation}

In this section, we motivate the use of explicit weak solutions to improve the efficiency of Monte Carlo methods, and present a weak solution to \eqref{eq:dFt} and \eqref{eq:dFt_P}. We the show how the solution can be used to construct an efficient simulation algorithm, which will be used to calculate the net liability defined in \eqref{eq:liabilityGMWB}. In the next section, the algorithm will be used to assess the efficiency of the VIX-linked fee in reducing the riskiness of the net loss.

\subsection{Expicit weak solutions used in Monte Carlo simulation}

Monte Carlo methods are popular in insurance pricing and risk assessment due to their generality and simplicity.  They apply to almost any model and can be implemented quickly on almost any computer system.  They also parallelize well.  Their most significant downfall is execution speed, especially on multi{-}dimensional or otherwise{-}complicated models.  There are two basic reasons for this heavy computational burden, the need to average out excess sampling noise and the need for accurate samples.  Still, research into reducing this burden is encouraging.

Excess sampling noise occurs due to the substitution of an empirical measure for a real probability measure. Traditionally, this is handled by using a lot of independent samples and relying on the law of large numbers.  However, the number of samples required for a decent level of accuracy on a stochastic model can be very large, creating great computational burden.  One good way to compensate for this issue is to use negatively correlated samples, which reduces the variance in estimates and allows one to use far fewer samples for the same accuracy.  One can quite easily introduce negative correlations into the driving processes in our models. It is however not clear how that would affect the correlations between samples. Therefore, more direct and model{-}independent methods are preferred.  In this vein, we  are interested in employing the quick simulation field methods of \cite{kouritzin2013random, kouritzin2014graph, kouritzin2017flexible}. These algorithms take marginal (conditional) distributions and desired correlations between samples and produce a joint distribution for all samples that is still easy to simulate in one pass. The joint distribution is one of many that has the given marginal distributions and correlations. They can be expressed in closed form. However, these methods still need to be generalized to continuous random variables and explored further.  Moreover, some level of negative correlation can be introduced in a different manner while improving sample accuracy as discussed below, and as done in \cite{kouritzin2017branching,kouritzin2017residual}.  Hence, it makes sense to investigate accurate samples first. 

Most often, for a general market model, one has to resort to producing approximate discretized model samples using the popular Euler or Milstein methods. Here, one draws new random variables and performs computations at each discretized time point.  The discretization has to be very fine to meet accuracy requirements or else the so{-}called discretization bias will be large.  Hence, the computational burden per sample is large unless alternative simulation methods are discovered.  The promising alternate simulation method used herein is based upon explicit weak solutions to our model. The motivating ideas as follow. Solutions to linear SDEs (like Ornstein{-}Uhlenbeck processes) can be trivially simulated, and time{-}dependent diffeomorphism of one SDE solution will often produce a solution of another possibly more{-}complicated SDE. Furthermore, weak solutions of complicated SDEs can sometimes be obtained by projecting a higher dimensional strong solution to an enlarged SDE onto lower dimensions.  Putting this together, we furthered the work in \cite{kouritzin2016path}, in which it is shown that explicit weak solution simulation methods compare favorably to Euler or Milstein methods, both in terms of speed and accuracy. In Section \ref{ssec:explicit}, we show that non{-}linear stochastic{-}volatility{-}jump SDEs of the form of \eqref{heston} can be also be represented, using explicit weak solutions, in a form that is very amenable to simulations. The details are below but we motivate this in the case $ w_{t}\equiv 0 $ now. First, when $\nu =\frac{n\kappa ^{2}}{4} $ for some $n\in \mathbb{N} $ we show in Section \ref{HestSimAlg} that $(S,V)$ have a solution of the form 
\begin{align*}
	S_{t} &= \phi _{t}\left(\int _{0}^{t}V_{s}^{\frac{1}{2}}dB_{s},\int _{0}^{t}V_{s}ds,V_{t}\right)\prod\limits _{s\le t}\left(1+\Delta X_{s}\right) \\
	V_{t} &=\sum\limits ^{n}_{i=1}\left(Y^{i}_{t}\right)^{2},
\end{align*}
where $\phi _{t} $ is some known function, $ \left\{Y^{i}\right\}_{i=1}^{n} $ are Ornstein{-}Uhlenbeck processes related to $ \beta  $ and $ \left(B,\beta ,X\right) $ are the driving processes of the SDE. Now, $ V_{t} $ can be simulated (exactly at any time) by Gaussian methods. $\int _{0}^{t}V_{s}^{\frac{1}{2}}dB_{s}$ is conditionally (on $ V $) zero{-}mean Gaussian and can be simulated (exactly and trivially) using its variance $ \int _{0}^{t}V_{s}ds $. $ \Delta X_{s} $ are just Poisson jumps.  Hence, everything can be simulated without any approximation or discretization other than using Simpson's rule or some other numerical integration method to evaluate the integral $ \int _{0}^{t}V_{s}ds $.

%

In the case,  $ \nu \ne \frac{n\kappa ^{2}}{4} $ for any $ n\in \mathbb{N} $, we change the measure to the case where  $ \nu =\frac{n\kappa ^{2}}{4} $ for some $ n\in \mathbb{N} $ using Girsanov's theorem. We show in Section \ref{HestSimAlg} that the  likelihood used for the change of measure can easily be simulated (conditionally on, i.e. after, $ V $) without approximation other than Simpson's rule.
	
There is still significant work to be done before the techniques in this paper could be called a method.  First, when $ \nu \ne \nu _{\kappa } $ the simulated samples are not the same as the desired samples and the likelihood $ L_{t} $ is used to correct back.  However, the sample likelihoods will generally diverge from each other and most samples become useless over long enough time.  The way to correct for this is to resample, interact or branch the samples when the likelihoods diverge too much.  Indeed, using ideas from sequential Monte Carlo one can introduce negative correlations into this resampling step, helping to reduce the ensemble (sampling error) variance while improving the individual particles. These branching techniques are incorporated in the algorithms presented in Appendix \ref{app:simulation_algorithms}.

%

\subsection{Explicit weak solutions for the VA account value}\label{ssec:explicit}

To price GMWB products, we recall the generalized SVJ model \eqref{eq:dFt} presented in Section \ref{ssec:VAaccount}:
\begin{align}
d
\bpm F_t \\ V_t \epm
= \bpm \mu F_t -\alpha V_t F_t - w_t \\ \nu-\varrho V_t \epm
dt +
\bpm \sqrt{1-\rho^2} F_t V^{\frac{1}{2}}_t & \rho F_t V^{\frac{1}{2}}_t  & F_{t^-}\\
0 & \kappa V^{\frac{1}{2}}_t & 0\epm
\bpm dB_t\\ d\beta_t \\ dX_t\epm,
\label{eq:HestonFt}
\end{align}
for $F_t>0$, and $dF_t = 0$ for $F_t \leq 0$.
In particular, the top equation drift has two added terms over the SVJ model introduced in \eqref{heston}.
We will show that explicit weak solutions also exist for the generalized model \eqref{eq:HestonFt}, and that
they can be used effectively to price GMWB products.

\subsubsection{Preliminary lemmas}
We first handle the additional drift term $w_t$.
\begin{lem}\label{lem:FtoH}
	Suppose that $\{(G_t,V_t),\ t\ge 0\}$ solves	
	\begin{equation}\label{eq:dG_t}
	d\!\left(\!\begin{array}{c} G_{t}\\ V_{t}\end{array}\!\right)
	=
	\left(\!\begin{array}{c}\mu G_{t}-\alpha V_t G_{t}\\\nu-\varrho  V_{t}\end{array}\!\right)dt
	+\left(\!\begin{array}{ccc}\sqrt{1-\rho^2} G_{t} V_{t}^{\frac{1}{2}}&\rho  G_{t} V_{t}^{\frac{1}{2}} & G_{t^-}\\0&\kappa  V_{t}^{\frac{1}{2}}&0\end{array}\!\right)\left(\!\begin{array}{c}dB_{t}\\d \beta_{t}\\dX_t\end{array}\!\right)
	\end{equation}
	subject to $G_0=1$ and $G_t>0$ for all $t$.
	Then, $\{(F_t,V_t),\ t\ge 0\}$ solves \eqref{eq:HestonFt}, where
	\begin{equation*}
	F_t=\left(F_0 G_t -\int_0^t\frac{G_t}{G_s}w_sds\right)
	\end{equation*}
	on $t < \tau_0$, and $F_t = 0$ on $t \geq \tau_0$, with $\tau_0= \inf\left\{t>0: F_t = 0\right\}$.
\end{lem}
Note that $G$ is a stochastic exponential and $P(G_t>0~ \forall t \in [0,T])=1$.

\proof
It follows by integration by parts that for $F_t > 0$,
\begin{equation*}
dF_t=(dG_t)\left[F_0-\int_0^t\frac{w_s}{G_s}ds\right]-w_t\, dt
\end{equation*}
and then \eqref{eq:HestonFt} follows by \eqref{eq:dG_t}. Note that $F_t$ is absorbed at $F_t=0$.
\qed

Next, for all $t \leq T$, we define the process $H$ by removing the jump part from $G$:
	\begin{equation}
	d\left(\begin{array}{c} H_{t}\\ V_{t}\end{array}\right)
	=
	\left(\begin{array}{c}\mu H_{t}-\alpha V_t H_{t}\\\nu-\varrho  V_{t}\end{array}\right)dt
	+\left(\begin{array}{cc}\sqrt{1-\rho^2} H_{t} V_{t}^{\frac{1}{2}}&\rho  H_{t} V_{t}^{\frac{1}{2}}\\0&\kappa  V_{t}^{\frac{1}{2}}\end{array}\right)\left(\begin{array}{c}dB_{t}\\d \beta_{t}\end{array}\right),
	\label{HestonKernal}
	\end{equation}
with $H_0 = G_0$. Note that $H$ is the modified Heston kernel used in \cite{cui2016VIX} in the context of guaranteed minimum accumulation benefits.	
Now we can write
\begin{align*}
	dG_{t} &= G_{t-}\left[(\mu -\alpha V_{t})dt+\sqrt{1-\rho ^{2}}V_{t}^{\frac{1}{2}}dB_t+\ \rho V_{t}^{\frac{1}{2}}d\beta_t+\ dX_t\right]\\
	dH_{t} &= H_{t}\left[(\mu -\alpha V_{t})dt+\sqrt{1-\rho ^{2}}V_{t}^{\frac{1}{2}}dB_t+\ \rho V_{t}^{\frac{1}{2}}d\beta_t\right],
\end{align*}
which means $G_t$ and $H_t$ are stochastic exponentials with (unique) solutions.  Hence, it is possible to obtain $G$ from $H$ via the following simple lemma.

\begin{lem}\label{lem:HtoG}
	Let $\Delta X_s = X_s - X_{s^-}$, and let $\{G_t\}_{t\leq T}$ and $\{H_t\}_{t\leq T}$ be defined by \eqref{eq:dG_t} and \eqref{HestonKernal}. Then,
	\begin{equation*}
	G_t = H_t\prod\limits _{s\le t}\left(1+\Delta X_s\right).
	\end{equation*}
\end{lem}

\proof
Let $M_t = \prod\limits _{s\le t}\left(1+\Delta X_s\right)$, which is a pure jump process, so the quadratic variation $[H,M]_t=0$ a.s. for all $t\leq T$, and $dM_t = M_{t^-}dX_t$. Hence, by integration by parts, 
\begin{align*}
	d(H_t M_t) &= H_{t^-} dM_t + M_{t^-} dH_t\\
	&= H_{t^-}M_{t^-}\left[dX_t + (\mu-\alpha V_t)dt + \sqrt{1-\rho^2}V^{\frac{1}{2}}_t dB_t + \rho V^{\frac{1}{2}}_t d\beta_t\right].
\end{align*}
The result follows since $G_0 = H_0 M_0$. 
\qed

For $t < \tau_0$, with $\tau_0$ as defined in \eqref{eq:tau0}, we can then write
\begin{align*}
F_{t}&= H_t \left(F_{0}\prod\limits _{s\le t}\left(1+\Delta X_{s}\right)- R_t\right)\vee 0,\\
R_t &= \int _{0}^{t}\frac{w_u}{H_{u}}\prod\limits _{u<s\le t}\left(1+\Delta X_{s}\right)du,
\end{align*}
where $ H $ is the explicit modified Heston model solution, which is presented below.

\subsubsection{Theoretical Result}
To solve $\{(H_t,V_t),\ t\ge 0\}$ weakly, we let
\begin{eqnarray*} 
\label{numuk}
n=\left\lfloor\frac{4\nu}{\kappa^2}+\frac12\right\rfloor\vee 1,\qquad \nu_\kappa=\frac{n\kappa^2}4,\qquad 
\mu_\kappa=\mu+\frac{\rho}{\kappa}\left(\nu_\kappa-\nu\right).
\end{eqnarray*}
Now, the modified Heston kernel \eqref{HestonKernal} has an explicit weak solution
either for all time if Condition (C) holds or else until the volatility drops too low.
\begin{theo} \label{Theorem2}
	Let $\varepsilon\in(0,1)$, $T>0$, 
	$\{W^1,...,W^n,B\}$ be independent standard Brownian motions with respect to the filtered probability space $(\Omega,\mathcal F,\{\mathcal F\}_{t\in [0,T]},Q)$, $V_0$ be given random variables with $V_0>\varepsilon$,
	\begin{eqnarray}\label{ExplicitStful}
	\!\!\!\!\!\!\!\!\!\!\!\! H_t&\!\!\!\!=&\!\!\!\! 
	\exp\!\bigg(\!\sqrt{1\!-\!\rho^2}\!\int_0^t \! V_s^\frac12 dB_s\!
	+\!\left[\mu\!-\frac{\nu\rho}\kappa\right]\! t\!
	+\!\left[\frac{\rho\varrho}\kappa-\frac{1}2-\alpha\right]
	\!\int_0^t \! V_s ds+\frac\rho\kappa ( V_t-\! V_0)\!
	\bigg)\ \ 
	\\\label{ExplicitVtful} \!\!\!\!\!\!\!\! V_t&\!\!\!\!=&\!\!\!\! \sum_{i=1}^n(Y_t^i)^2,
	\ \ \eta_\varepsilon=\inf\left\{t: V_t\le\varepsilon\right\}\ \text{ and}\\
	\!\!\!\!\!\!\!\!\label{L2way}
	L_t&\!\!\!=&\!\!\!\!\exp\left\{ \frac{\nu-\nu_\kappa}{\kappa^2} \left[\ln( V_t)-\ln( V_0)
	+\int_0^t\frac{\kappa^2-\nu_\kappa-\nu}{2 V_s}+\varrho\, ds
	\right] \right\},
	\end{eqnarray}
	where $Y_t^i=\frac{\kappa}2\int_0^t e^{-\frac{\varrho}2 (t-u)}dW^i_u+e^{-\frac{\varrho}2t}Y^i_0$ for $i=1,2,...,n$.
	Define 
	\begin{eqnarray}
	\label{Bbetadef}
	\!\!\!\!\!\!\!\beta_t&\!\!=&\!\!\sum_{i=1}^n \int_0^t\frac{Y_u^i}{\sqrt{\sum_{j=1}^n(Y_u^j)^2}}dW^i_u+\int_0^{t\wedge\eta_\varepsilon} \frac{\nu-\nu_\kappa}{\kappa  V_s^\frac12}ds,\ \ \text{and}\\\label{Phatdef}
	\!\!\!\!\!\!\! \widehat Q(A)&\!\!=&\!\!E[1_A  L_{T\wedge\eta_\varepsilon}]\ \ \forall A\in \mathcal F_T.
	\end{eqnarray}
	Then, $\eta_\varepsilon$ is a stopping time and $ L_{t\wedge\eta_\varepsilon}$ is a
	$L^r$-martingale with respect to $Q$ for any $r>0$.
	Moreover, $(B,\, \beta)$ are independent standard Brownian motions and
	\begin{eqnarray}\label{WeakHestonPtilde}
	d\!\left(\!\begin{array}{c} H_{t}\\ V_{t}\end{array}\!\right)
	=
	\left\{\!\!\begin{array}{cc}
	\left(\!\begin{array}{c}(\mu -\alpha V_t) H_t\\\nu-\varrho  V_{t}\end{array}\!\right)dt+\left(\!\begin{array}{cc}\sqrt{1-\rho^2} H_{t} V_{t}^{\frac{1}{2}}&\rho  H_{t} V_{t}^{\frac{1}{2}}\\0&\kappa  V_{t}^{\frac{1}{2}}\end{array}\!\right)\left(\!\begin{array}{c}dB_{t}\\d \beta_{t}\end{array}\!\right),
	& t\le \eta_\varepsilon\\
	\left(\!\begin{array}{c}(\mu_\kappa -\alpha V_t) H_t\\\nu_\kappa-\varrho  V_{t}\end{array}\!\right)dt+\left(\!\begin{array}{cc}\sqrt{1-\rho^2} H_{t} V_{t}^{\frac{1}{2}}&\rho  H_{t} V_{t}^{\frac{1}{2}}\\0&\kappa  V_{t}^{\frac{1}{2}}\end{array}\!\right)\left(\!\begin{array}{c}dB_{t}\\d \beta_{t}\end{array}\!\right),
	& t> \eta_\varepsilon
	\end{array}\right.		
	\end{eqnarray}
	on $[0,T]$ with respect to $\widehat Q$.
\end{theo} 

\proof
We just provide \emph{guess-and-check} style proof here and put the changes required
to the motivated proof of \cite{kouritzin2016path} in the Appendix.
$\{Y^i\}_{i=1}^n$
are independent Ornstein-Uhlenbeck processes satisfying
\begin{equation*}
dY^i_t =-\frac\varrho2 Y^i_t dt+\frac\kappa2 dW^i_t	
\end{equation*}
so 
it follows by It\^{o}'s formula
that $ V_t=\sum_{i=1}^n(Y_t^i)^2$ satisfies 
\begin{equation}\label{Vequation}
dV_t=2\sum_{i=1}^nY^i_t\ dY^i_t+\frac{n\kappa^2}4dt=(\nu_\kappa-\varrho  V_{t})dt+\kappa  V_{t}^{\frac{1}{2}}d\widehat \beta_t	
\end{equation}
with $Q$-Brownian motion
\begin{equation}\label{Bhatdefine}
\widehat \beta_{t}=\sum_{i=1}^n \int_0^t\frac{Y_u^i}{\sqrt{\sum_{j=1}^n(Y_u^j)^2}}dW^i_u. 
\end{equation}
Moreover, using the fact $\mu\!-\frac{\nu\rho}\kappa=\mu_\kappa\!-\frac{\nu_\kappa\rho}\kappa$ and It\^{o}'s formula, one finds
the right hand side of \eqref{ExplicitStful} satisfies
\begin{eqnarray}\label{Tdiff}
dH_t
&=&H_t\left[\sqrt{1\!-\!\rho^2} V_t^\frac12 dB_t\!
+\!\left[\mu_\kappa\!-\frac{\nu_\kappa\rho}\kappa\right]\! dt\!
+\!\left[\frac{\rho\varrho}\kappa-\frac{1}2-\alpha\right]
\! V_t dt+\frac\rho\kappa d V_t\right]
\\\nonumber
&+&\frac{H_t}2\left[(1\!-\!\rho^2) V_t dt\!
+\frac{\rho^2}{\kappa^2} \kappa^2 V_tdt\right]
\\\nonumber
&=&
(\mu_\kappa -\alpha V_t) H_tdt+\sqrt{1-\rho^2} H_{t} V_{t}^{\frac{1}{2}}dB_{t}+\rho  H_{t} V_{t}^{\frac{1}{2}}d \widehat\beta_{t}	
\end{eqnarray}
by \eqref{Vequation} and the fact $B$ and $\widehat \beta$ are independent.
We have recovered the lower equations in \eqref{WeakHestonPtilde}.
The upper equations and the switch are obtained by a Girsanov measure change $Q\rightarrow\widehat Q$, defined in \eqref{Phatdef},
which makes $\beta$, defined in \eqref{Bbetadef}, a Brownian motion independent of $B$.
With this in mind, one finds by It\^{o}'s formula and \eqref{Vequation} that
\begin{equation*}
\ln( V_t)-\ln( V_0)=\int_0^t \frac{\nu_\kappa-\varrho  V_s}{ V_s}ds +\int_0^t \frac{\kappa}{ V_s^\frac12}d\widehat \beta_s-\frac12\int_0^t \frac{\kappa^2}{ V_s}ds
\end{equation*}
so, using \eqref{numuk}, \eqref{L2way} is equivalent to
\begin{equation}\label{L1way}
L_t=\exp\left\{\int_0^t \frac{\nu-\nu_\kappa}{\kappa  V_s^\frac12}d\widehat \beta_s
-\frac12\int_0^t \frac{|\nu-\nu_\kappa|^2}{\kappa^2  V_s}ds
\right\}.
\end{equation}
Then, following \cite{kouritzin2016path}, one can easily show that 
\begin{eqnarray*}
	&\!\!\!\!&\!\!\!\!E\!\left[ L^{\eta_\varepsilon}_T\left(f( H_{t_{n+1}}, V_{t_{n+1}})-f( H_{t_{n}}, V_{t_{n}})
	-\int_{t_{n}}^{t_{n+1}}\!\!A_uf( H_u, V_u)du\right)
	\prod_{k=1}^nh_k( H_{t_k}, V_{t_k})\right]=0
\end{eqnarray*}
for all $0\le t_1< t_2<\cdots<t_n<t_{n+1}$, $f\in \mathcal S(\mathbb R^2)$ (the rapidly decreasing functions) and $h_1,...,h_n\in B(\mathbb R^2)$ (the bounded, measurables),
where
\begin{eqnarray*}
A_uf(\tau,v)&\!\!\!=&\!\!\![(\mu-\alpha v) \tau \partial_\tau f(\tau,v)+(\nu-\varrho v) \partial_vf(\tau,v)]1_{[0,\eta_\varepsilon]}(u)
\\\nonumber
&\!\!\!+&\!\!\![(\mu_\kappa -\alpha v)\tau \partial_\tau f(\tau,v)+(\nu_\kappa-\varrho v) \partial_vf(\tau,v)]1_{[\eta_\varepsilon,T]}(u)\\\nonumber
&\!\!\!+&\!\!\!\frac12 \tau^2 v \partial^2_\tau f(\tau,v)+\rho\kappa\partial_v\partial_\tau f(\tau,v)+\frac{\kappa^2}2\partial^2_vf(\tau,v).
\end{eqnarray*}
(We are using $\tau,v$ as state varibles for $H,V$ respectively.)
Now, it follows by the argument on page 174 of \citet{Ethier/Kurtz:1986} that 
$(H,V)$ satisfies the $A_u$-martingale problem with respect to $\widehat Q$.
\qed

The solution \eqref{ExplicitStful}-\eqref{Bbetadef} is valid for \emph{any} $\{Y_0^i\}_{i=1}^n$ such that $\sum_{i=1}^n(Y_0^i)^2= V_0$.
When Condition (C) is true, $L_t= 1$ and both $L_t$ and $\eta_\varepsilon$ are redundant.
The solution is constructed with respect to the original risk-neutral probability $Q$ and satisfies
the upper equation in (\ref{WeakHestonPtilde}) until $T$.
Otherwise, the solution satisfies the desired Heston model until $\eta_\varepsilon$, 
i.e.\ until the volatility drops
too low (or we hit the final `simulation time' $T$) and then falls back to the
closest explicit alternative given in the lower part of (\ref{WeakHestonPtilde}).
In this case the solution is with respect to
a manufactured probability $\widehat Q$.

\subsection{Weighted and explicit simulation of the VA account}\label{HestSimAlg}

Now, we use Theorem \ref{Theorem2} to produce an efficient algorithm to simulate the value of the VA account and the volatility process. The algorithm will then be used to price the GMWB. The algorithm can also serve to price other types of guarantees on the VA account described by \eqref{eq:dFt}, such as the GMAB analyzed in \cite{cui2016VIX}, or to model a VA account with flexible premium.

Defining constants 
\begin{equation*}
a=\sqrt{1-\rho^2},\ b=\mu-\frac{\nu\rho}{\kappa},\ c=\frac{\rho\varrho}{\kappa}-\frac{1}{2}-\alpha,\
d= \frac{\rho}{\kappa},\ e=\frac{\nu-\nu_\kappa}{\kappa^2},\ f=e\frac{\kappa^2-\nu-\nu_\kappa}{2}, 
\end{equation*}
we find that \eqref{ExplicitStful} and \eqref{L2way} can be rewritten as
\begin{eqnarray}\label{Ttfulabcd}
\!\!\!\!\!\!\!\!\!\!\!\! H_{t_k}&\!\!\!\!=&\!\!\!\!  H_{t_{k-1}}
\exp\!\bigg(\!a\int_{t_{k-1}}^{t_{k}} \! V_s^\frac12 dB_s\!
+b(t_k - t_{k-1}) +c \int_{t_{k-1}}^{t_{k}} \! V_s ds+d\ ( V_{t_k}-\! V_{t_{k-1}})\!\bigg)
\\
\!\!\!\!\!\!\!\!\label{Lef}
L_{t_{k}}&\!\!\!=&\!\!\!\! L_{t_{k-1}}\exp\left\{ e \left(\ln\left(\frac{ V_{t_k}}{ V_{t_{k-1}}}\right)+\varrho(t_k - t_{k-1})\right) 
+f\int_{t_{k-1}}^{t_{k}}\frac{1}{ V_s}\,ds
\right\}.
\end{eqnarray}
The stochastic integral in \eqref{Ttfulabcd} is conditionally (given $ V$) Gaussian since $ V$ 
and $B$ are independent so simulation is just a centered normal random variable with
variance $a^2\int_{t_{k-1}}^{t_{k}} \! V_s ds$.
Even the weight \eqref{Lef} avoids stochastic integrals.
There are a number of choices for the two deterministic integrals 
to be computed. For example, using the trapezoidal rule, we have
\begin{equation*}
\int_{t_{k-1}}^{t_{k}} \! V_s ds\approx \frac1{2M}\left\{ V_{t_{k-1}}+ V_{t_k}+2\sum_{l=1}^{M-1} V_{t_k-\frac{l}{M}h}\right\},
\end{equation*} 
with $h=t_k-t_{k-1}$, and a similar formula for $\int_{t_{k-1}}^{t_{k}} \!\frac1{ V_s} ds$.

The Weighted Simulation algorithm is given in details in Appendix \ref{app:simulation_algorithms}. Note that when ${\nu=\frac{n\kappa^2}4}$, the explicit solutions do not require the weights $L_t$. In this case, $L_t$ and $\eta_\varepsilon$ should not be simulated. This reduced algorithm is then called the \emph{Explicit Simulation} algorithm, and the general one (presented in Appendix \ref{app:simulation_algorithms}) stemming from Theorem \ref{Theorem2} is the \emph{Weighted Simulation} algorithm.

When $\nu - \nu_\kappa \neq 0$, the weighted simulation algorithm can also be improved by random resampling, or branching, of the simulated particles. At each time step, particles are branched individually, if necessary. The probability of re-sampling from a given path is based on the associated likelihood $L_t$. Different branching algorithms are presented in \cite{kouritzin2017branching}. The algorithms were originally developed in the context of sequential Monte Carlo methods in \cite{kouritzin2017residual} and analyzed in \cite{kouritzin2017convergence}. Here, for optimal performance, we use the \textit{Effective Particle Branching} algorithm, which we present in details in Subroutine \ref{algo:BranchingOneStep} in Appendix \ref{app:simulation_algorithms}.

\begin{re}
To understand the need to stop (at $\eta_\varepsilon$) before the volatility gets too small in Algorithm \ref{algo:WeightedOneStep}, 
we consider the situation where the volatility $V_t^\frac12=0$.
Then, the (closest explicit and general) volatility equations become deterministic
\begin{equation*}
dV_t=\nu_\kappa ~dt,\ \ \ \ dV_t=\nu ~dt,
\end{equation*}
and it is obvious which solution one has.
This makes model distributions singular to each other when $\nu_\kappa\ne \nu$.
\end{re}

The weighted or branching simulation algorithm can be used to calculate the initial net liability of the GMWB, $\Pi^W(0,F_0,V_0)$ given by \eqref{eq:liabilityGMWB}. The variable annuity account value is obtained with the help of Lemmas \ref{lem:FtoH} and \ref{lem:HtoG}. We define the additional variables $C_t$ and $W_t$ by
\begin{equation}
C_t = \int_0^{t \wedge \tau} e^{-ru}c_u F_u ~ du
\label{eq:Ct}
\end{equation}
and
\begin{equation}
W_t = \int_{\tau}^{t}e^{-ru} w_u ~ du
\label{eq:Wt}
\end{equation}
to track the discounted total fee paid up to $t$ and the discounted total GMWB payoff paid up to $t$, respectively.  The net liability is then obtained using a weighted Monte Carlo estimate. After all values $\{(C_T^j,W_T^j,L_T^j)\}_{j=1}^N$ have been simulated, the Monte Carlo estimate of the net liability is given by
\begin{equation*}
\widehat{\Pi}^W(0,F_0,V_0) = \frac{1}{N}\sum_{j=1}^{N} (W^j_T - C^j_T) L^j_T.
\end{equation*}
The details of this procedure are presented in Appendix \ref{app:simulation_algorithms}.

\section{Numerical results}\label{sec:4_numerical_results}

In this section, we use the simulation algorithm introduced in Section \ref{HestSimAlg} to analyze the effectiveness of the VIX-linked fee in reducing the sensitivity of the net liability and the net initial loss of a GMWB. We first explore the fair fee in the context of the VIX-linked fee structure. Then, we assess the impact of the fee structure on the net liability and the distribution of the net loss.

\subsection{Market parameters}

While the calculations of the fair fee and the net liability only require a calibration under the risk-neutral measure $Q$, the distribution of the net loss should be obtained under the real-world measure $P$. Calibrations of the SVJ model to both the $P$- and $Q$-measures are rare in the literature, and even more so when we require that they use relatively recent data. When they exist, they are not always appropriate for our purpose. For example, the parameters of \cite{duan2010jump} result in a stochastic volatility process that is no longer mean-reverting under the $Q$-measure. Since the simultaneous calibration of the SVJ model to both measures is out of the scope of this paper, we have chosen to use the risk-neutral parameters obtained by \cite{kokholm2015joint}. For the calculation of the distribution of the net loss, the $P$-measure parameters are obtained by choosing reasonable values for the different risk premiums.

The parameters we use throughout the numerical examples, unless otherwise indicated, are presented in Table \ref{tab:Parameters}. They were obtained by \cite{kokholm2015joint} using SPX option data on May 16, 2012. Note that we modified the value of $\nu$ (that is, $\kappa\theta$ in the original paper notation) from $0.1773$ to $0.18$, to ensure that Condition (C) holds. This modification speeds up our calculation without having a significant impact on our results. In addition, we set the risk-free rate to $r=0.02$. 

\begin{table}[h!]
	\caption{SVJ model parameters, modified from \cite{kokholm2015joint} (calibration to SPX option data on May 26,2012)}
	\label{tab:Parameters}
	\centering
	\begin{tabular}{cccccccc}
		\hline
		$\nu$ & $\varrho$ & $\kappa$ & $V_0$ & $\rho$ & $\lambda$ & $\delta$ & $\chi$  \\
		0.18 & 2.86 & 0.6 & 0.04 & -0.96 & 0.21 & -0.1252 & 0.18\\
		\hline
	\end{tabular}
	
\end{table}

\subsection{Fair fee rate}\label{ssec:ExFfee}

To illustrate the effects of the VIX-linked fee structure, throughout the section we consider a VA contract with GMWB rider, with initial premium $F_0=100$. We assume that the policyholder can withdraw up to $7\%$ of the initial account value per year, and that the contract is priced assuming that the maximum amount is withdrawn each year, continuously (that is, $w_t = 7$).

We first study the effect of different fee structures $(\bar{c},m)$ on the expected discounted total fee $C_T$ and on the expected discounted payout $W_T$, as defined in \eqref{eq:Ct} and \eqref{eq:Wt}. We consider $m \in \{0,0.1,0.2,0.3\}$ and $\bar{c} \in [0.005,0.035]$. For each fee structure $(\bar{c},m)$, $C_T$ and $W_T$ are calculated using the explicit simulation method introduced in Section \ref{HestSimAlg}, with $N=2\times10^5$ simulated paths and a time step $h=1/250$. 

\begin{figure}[!ht]
	\centering
	\subfigure[Expected discounted fees] {\includegraphics[width=7.5cm]{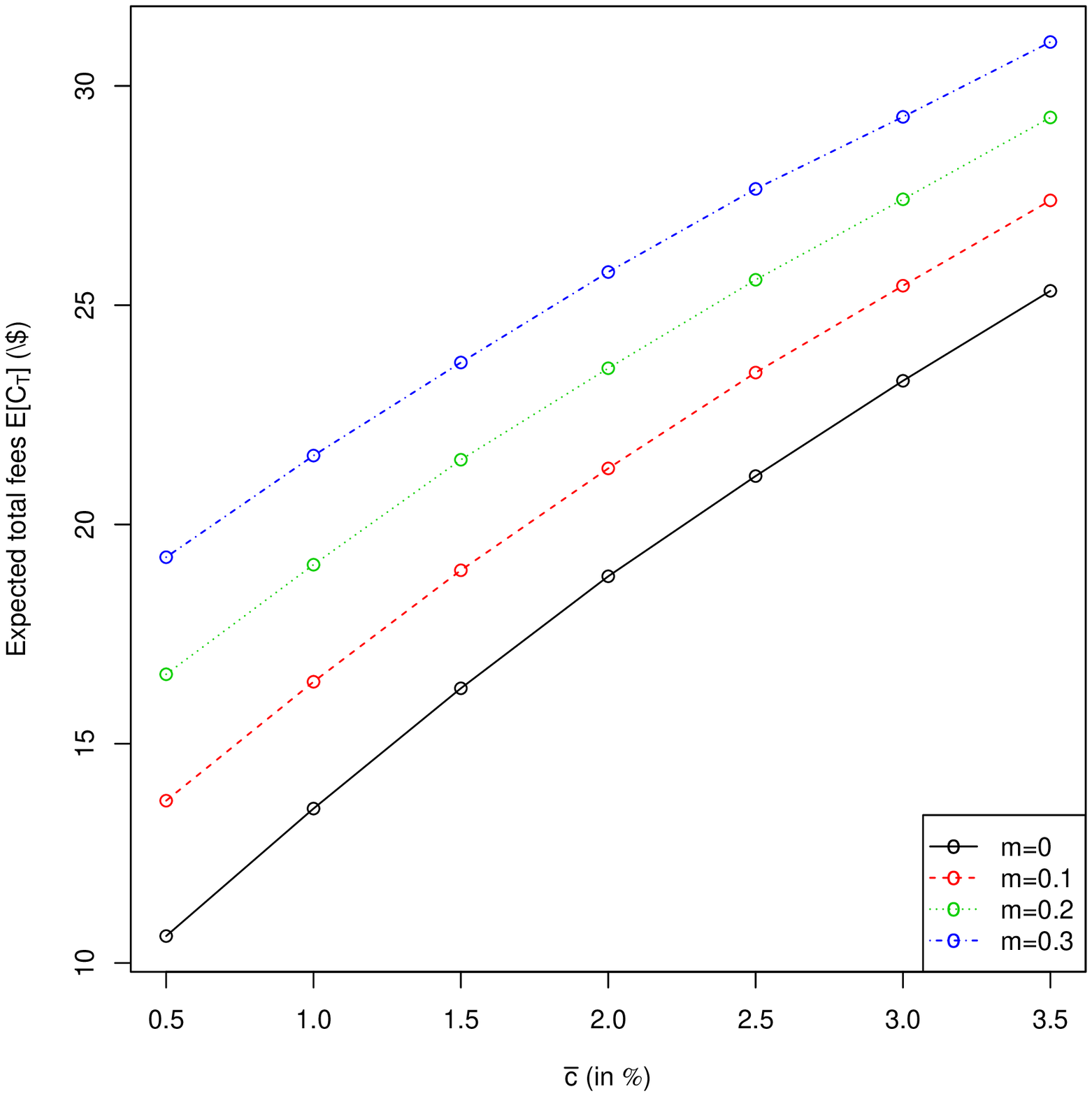}\label{fig:Ct}}\qquad
	\subfigure[Expected discounted payoff] {\includegraphics[width=7.5cm]{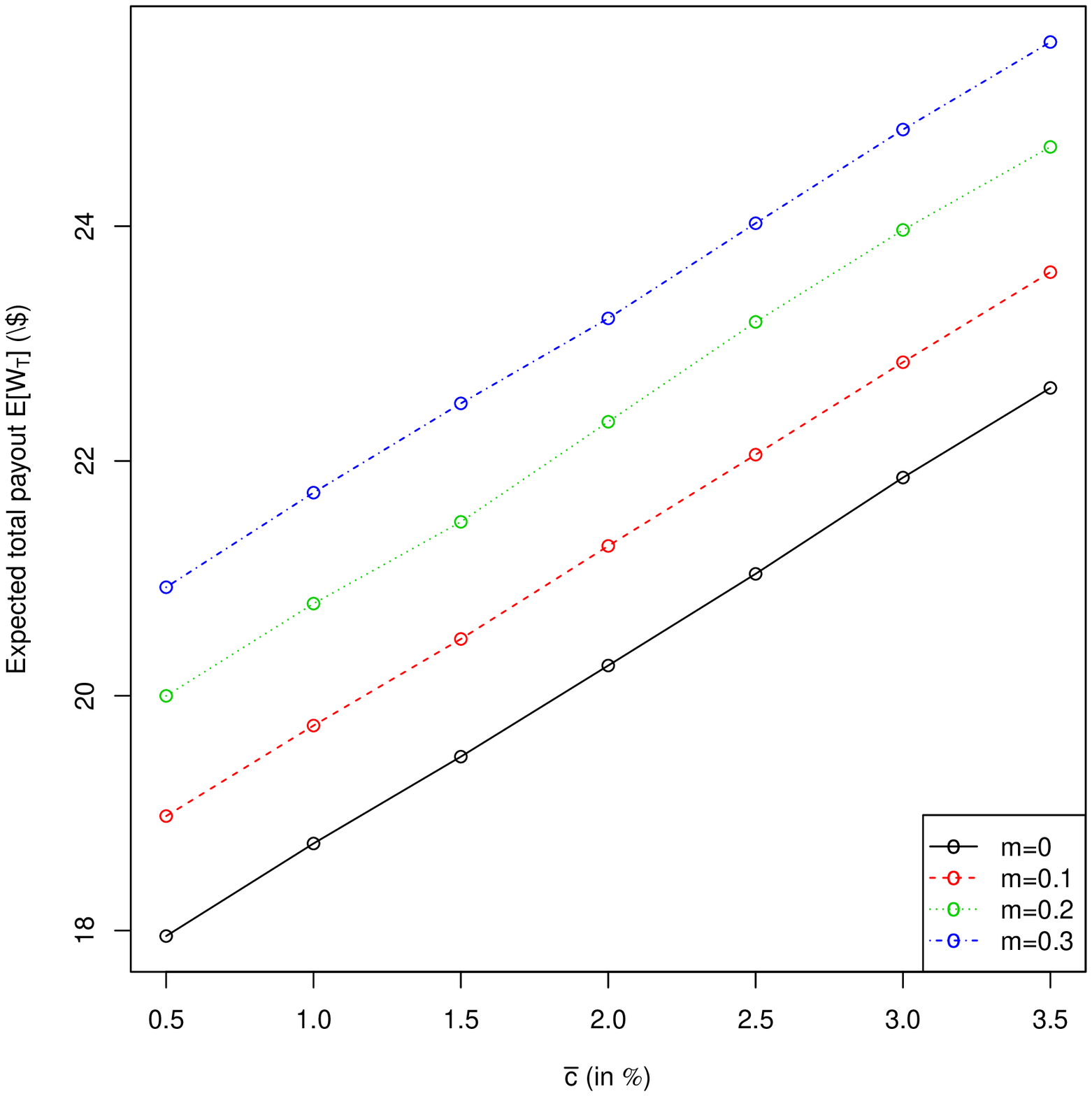}\label{fig:Wt}}
	\caption{Expected discounted fees $C_T$ and expected discounted payout $W_T$ as a function of the base fee $\bar{c}$ for various levels of the multiplier $m$.}
	\label{fig:CW_plots}
\end{figure}

Figure \ref{fig:CW_plots} presents the expected discounted total fee and payout, as defined in \eqref{eq:Ct} and \eqref{eq:Wt} as a function of $\bar{c}$, for different levels of $m$. For a fixed $m$, both the total fee and the payout are increasing in $\bar{c}$, for all four values of $m$ studied. It is clear that the total fee should increase in $\bar{c}$, since the total fee rate is the sum of $\bar{c}$ and another positive value, the square of the VIX multiplied by $m$. The total payout also increases with $\bar{c}$, since the increased fee rate directly reduces the net return on the VA account, so that a higher fee rate leads to a faster depletion of the fund. Hence, on average, when $\bar{c}$ is increased, the first hitting time of $0$ for $F_t$ will be smaller, and the guarantee will be paid out for a longer period of time, resulting in a higher average total payout.

As explained in Section \ref{ssec:pricingVAs}, a fair fee structure $(\bar{c}^*,m^*)$ is such that, under the risk-neutral measure, the expected payout from the insurer is equal to the expected fee income. Using fair fee rates to analyze the impact of the VIX-linked fee allows us to compare the different types of fee structure. For example, by letting $\bar{c}$ be such that the net liability at inception of the contract is null, we can assess the effect of setting the multiplier $m$ at different levels. By ensuring that all fee structures considered are fair, we remove the impact of the initial liability on the results. We will also relax the assumption of a fair fee structure to analyze the effect of the VIX-linked fee on underpriced products in a later section.

To obtain fair fee structures $(\bar{c}^*,m^*)$, we fix the multiplier $m$ to $0$, $0.1$, $0.2$ and $0.3$. For each value of $m$, we solve for the corresponding $\bar{c}$ that satisfies \eqref{eq:ffee_condition}. Note that when $m=0$, the fee rate is constant, rather than dependent on the value of the VIX index. Using $m=0$ allows us to compare the VIX-linked fee structure to the fixed fee one, which is typically used in the industry. For a given multiplier $m$, the net liability is calculated using the explicit simulation method introduced in Section \ref{HestSimAlg}, with $N=200000$ simulated paths and a time step $h=1/250$.

\begin{figure}[!ht]
	\centering
	\includegraphics[width=7.5cm]{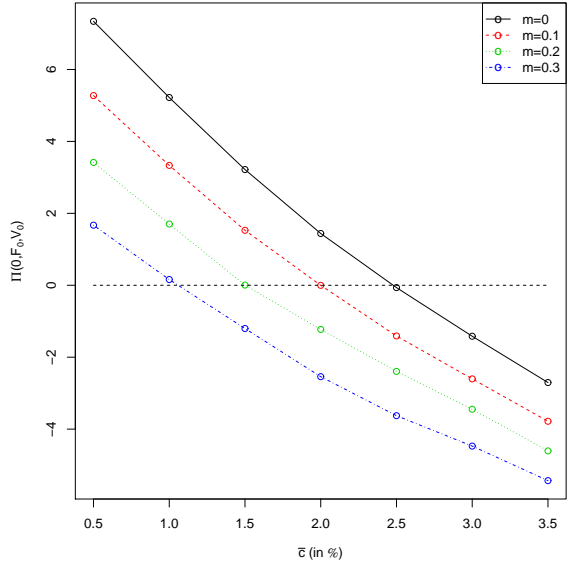}
	\caption{Net liability as a function of the base fee $\bar{c}$ for various levels of the multiplier $m$.}
	\label{fig:fair_fee_plots}
\end{figure}

On the interval $\bar{c} \in \[0.005,0.035\]$, Figure \ref{fig:fair_fee_plots} shows that the net liability is strictly decreasing in $\bar{c}$, and that for all four values of $m$ considered, it is possible to find $\bar{c}$ such that the fee structure is fair. Similarly to the findings of \cite{cui2016VIX} for the GMAB, a higher multiplier $m$ results in a lower fair $\bar{c}^*$. Indeed, when the VIX-linked component of the fee rate is higher, the remaining constant component can be lower, in order to cover the same financial guarantee. 

The fair fee rates used in the rest of this section are presented in Table \ref{tab:FairFee}.

\begin{table}[h!]
	\caption{Fair fee rates for a GMWB contract, $F_0=100$, $w_t=7$, $q=0.0075$.}
	\label{tab:FairFee}
	\centering
	\begin{tabular}{lcccc}
		\hline
		$m^*$ & 0 & 0.1 & 0.2 & 0.3 \\
		$\bar{c}^*$ (in \%) & 2.4650 & 1.9859 & 1.5275 & 1.0300 \\
		\hline
	\end{tabular}
	
\end{table}

\subsection{Net liability}

As explained in \cite{CBOE}, one of the goals of the VIX-linked fee is to reduce the sensitivity of the net liability to changes in market volatility. Indeed, in periods of high volatility, the expected value of the future payout tends to rise. When fees are set at a constant rate, the expected fee income remains at the same level, and can even fall, because of the leverage effect. Therefore, an increase in the current volatility will result in increased net liabilities. Such a variation leads to higher hedging costs and may also impact capital requirements. 

In the case of the GMAB, \cite{cui2016VIX} show that VIX-linked fees help reduce the effect of high volatility on the net liability. In the case of the GMWB, it is reasonable to expect a similar impact. If the volatility increases before the account is depleted, we can expect lower returns, which will speed up the depletion of the account and increase the expected value of the future payout. With a constant fee rate, such a rise in the volatility pushes the net liability up. However, with a VIX-linked fee structure, the expected future income also rises, thus reducing the overall effect of the volatility on the net liability. 

To illustrate the impact of the fee structure, we consider the same contract as in Section \ref{ssec:ExFfee} and calculate the net liability, defined in \eqref{eq:net_liability_GMWB} for different values of $V_0$. We are indeed interested in the impact of a temporary increase in market volatility on the value of the net liability. Reducing the sensitivity of the net liability to transitory changes in market volatility will help smooth out the value of the payoff to hedge, which should in turn improve the efficiency of the insurer's hedging strategy.

\begin{figure}[!ht]
	\centering
	\subfigure[Fairly priced guarantees]
	{\includegraphics[width=7.5cm]{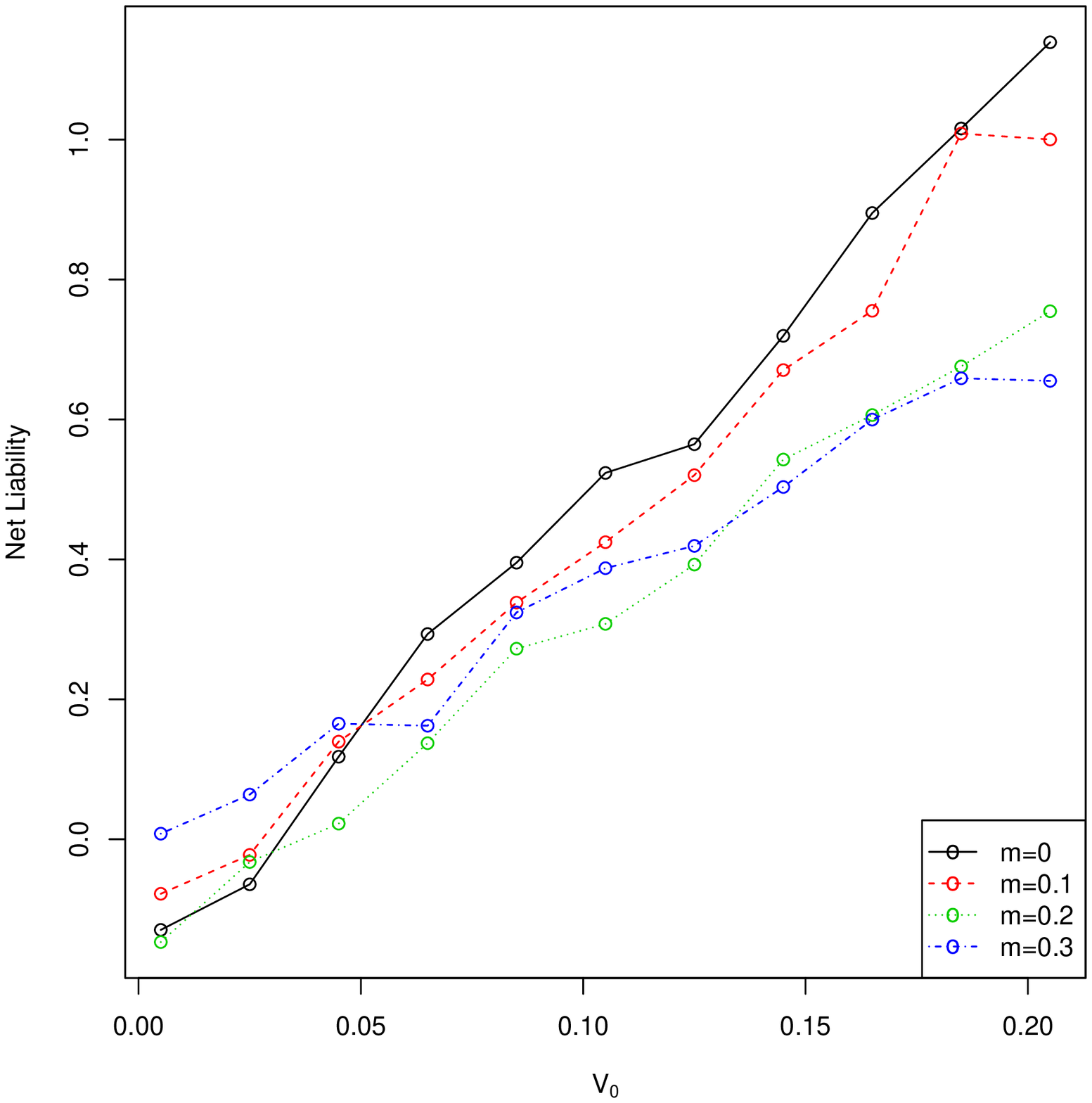}\label{fig:NL_sens_SVJ_fair}}
	\qquad
	\subfigure[Underpriced guarantees]
	{\includegraphics[width=7.5cm]{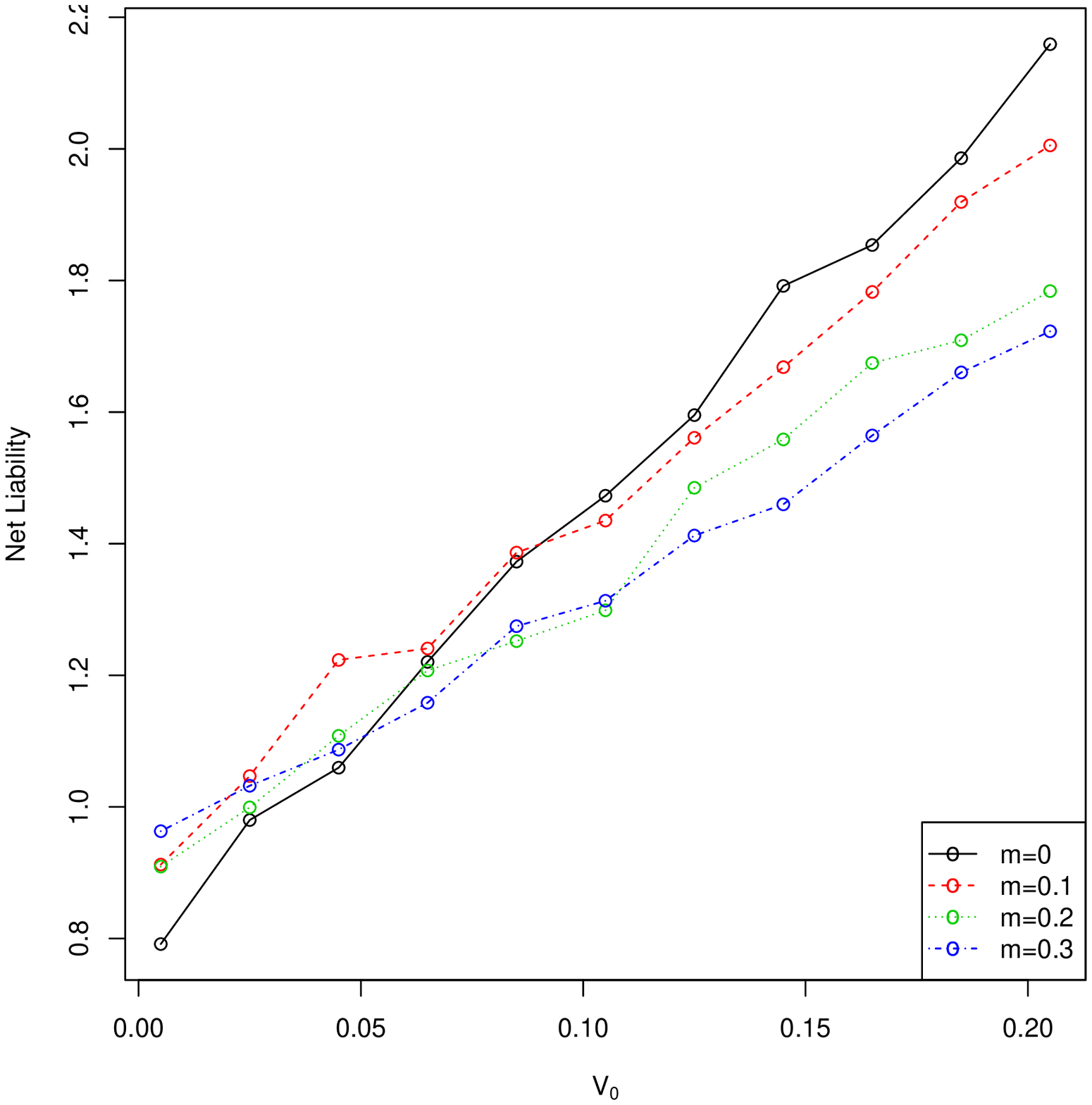}\label{fig:NL_sens_SVJ_low_cbar}}	
	\caption{Net liability at $t=0$ as a function of $V_0$.}\label{fig:NL_sens_SVJ}
\end{figure}

In Figure \ref{fig:NL_sens_SVJ}, we present the net liability at inception of the contract for different values of $V_0$. The results presented in this section were obtained using $N = 5 \times 10^5$. In Figure \ref{fig:NL_sens_SVJ_fair}, the results are obtained assuming that the contracts are priced fairly, using the fees given in Table \ref{tab:FairFee}. The results in Figure \ref{fig:NL_sens_SVJ_low_cbar} reflect an underpriced contract\footnote{Underpricing could be due, for example, to the use of a different pricing paradigm, or to a re-calibration of the pricing model.}; for $m \in \{0,0.1,0.2,0.3\}$, the corresponding base fee $\bar{c}$ is set such that under the initial parameters (from Table \ref{tab:Parameters}), the net liability at inception is equal to 1. Both graphs in Figure \ref{fig:NL_sens_SVJ} confirm that as $V_0$ increases, the net liability is less affected when the fee is linked to the VIX index ($m>0$). The effect is more noticeable for higher values of $m$. In other words, having a larger portion of the fee rate linked to the VIX helps reduce the impact of an increase in the current volatility. On the other hand, the VIX-linked fee seems to have a reduced effect on the net liability of an underpriced contract.

\begin{figure}[!ht]
	\centering
	\includegraphics[width=7.5cm]{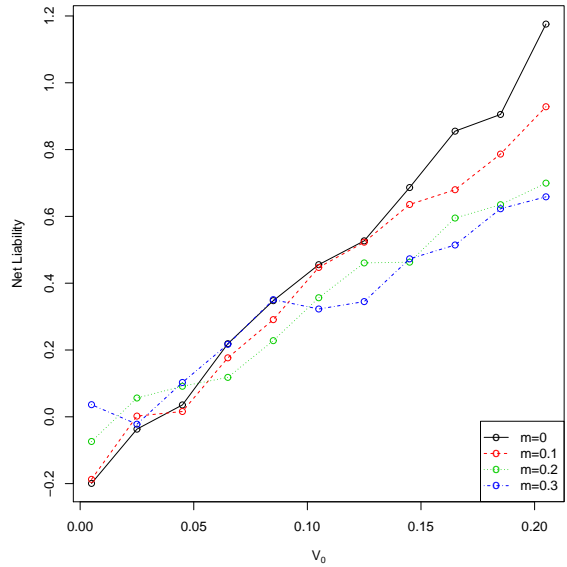}
	\caption{Net liability at $t=0$ as a function of $V_0$, $\rho = -0.3$.}\label{fig:NL_sens_SVJ_low_rho}
\end{figure}

The effectiveness of the VIX-linked fee in making the net liability more robust to changes in the volatility depends on the negative correlation between equity returns and the VIX. Therefore, it is relevant to assess the impact of the VIX-linked fee when this correlation is weaker. Figure \ref{fig:NL_sens_SVJ_low_rho} presents the net liability for different \emph{fair} fee structures ($m \in \{0,0.1,0.2,0.3\}$) when $\rho = -0.3$, and shows that the VIX-linked fee can still reduce the sensitivity of the net liability to changes in the spot volatility.

\subsubsection{Varying withdrawal rate}
Monte Carlo simulations allow for great flexibility in modeling. In particular, in our framework, the withdrawal rate does not need to be constant. In this section, we assess the impact of a VIX-linked fee on the net liability of a GMWB policy, given that it is calculated assuming time-varying withdrawals. We consider the same contract as in the previous section (that is, we use $F_0 = 100$) and the following withdrawal assumptions (also summarized in Table \ref{tab:withdrawals2}):
\begin{enumerate}
	\item
	\textit{Deferred withdrawals:} The liability is calculated under the assumption that no withdrawals will occur during the first 5 years of the contract, after which withdrawals will be made at an annual rate of $10\%$ of the original premium. That is, we set $w_t = 0$ for $0 \leq t < 5$ and $w_t = 10$ for $5 \leq t < 15$.
	\item
	\textit{Increasing withdrawals:} Under this assumption, the annual withdrawal rate is $5\%$ of the initial premium for the first two years, after which it increases to $6\%$ in year three, and subsequently by $1\%$ every three year. It therefore reaches $9\%$ at the beginning of year 12, and the guarantee is entirely withdrawn after 14 years.
	\item
	\textit{Decreasing withdrawals:} This withdrawal pattern assumes that during the first three years, withdrawals are made at a rate of $9\%$ of the initial premium, and then decrease by $1\%$ every three years until the guarantee runs out, which happens after 14 years. In the last two years of the contract, withdrawals are thus assumed to be made at an annual rate of $5\%$ of the initial premium.
\end{enumerate}


\begin{table}[h!]
	\caption{Withdrawal rate assumptions}
	\label{tab:withdrawals2}
	\centering
	\begin{tabular}{lccccccccccccccc}
		\hline
		\textbf{Year} & \textbf{1} & \textbf{2} & \textbf{3} & \textbf{4} & \textbf{5} & \textbf{6} & \textbf{7} & \textbf{8} & \textbf{9} & \textbf{10} & \textbf{11} & \textbf{12} & \textbf{13} & \textbf{14} & \textbf{15} \\
		$w_t$ (Deferred) & 0 & 0 & 0 & 0 & 0 & 10 & 10 & 10 & 10 & 10 & 10 & 10 & 10 &10 & 10 \\
		$w_t$ (Increasing) & 5 & 5 & 6 & 6 & 6 & 7 & 7 & 7 & 8 & 8 & 8 & 9 & 9 & 9 & 0 \\
		$w_t$ (Decreasing) & 9 & 9 & 9 & 8 & 8 & 8 & 7 & 7 & 7 & 6 & 6 & 6 & 5 & 5 & 0 \\
		\hline
	\end{tabular}
	
\end{table}

These withdrawal assumptions were picked to assess the influence of the withdrawal pattern on the fair VIX-linked fee structure and the resulting liability. The first assumption, in which withdrawals are deferred by a number of years, could represent the situation where a policyholder buys the policy with the intent of starting the withdrawals at a later date. In reality, in that case, the guaranteed amount available for withdrawal (usually the amount of the initial premium) would typically increase during the first years, in which no withdrawals occur. Increasing and decreasing withdrawals are not typical assumptions for pricing purposes. They are used here to illustrate the flexibility of our framework, and to test the robustness of our results on the VIX-linked fee to different withdrawal patterns. Our case studies are simplified versions of reality. For example, we ignore the penalties usually applied to withdrawals that are above the maximum annual rate allowed.

\begin{figure}
		\centering
		\subfigure[Deferred withdrawals]
		{\includegraphics[width=7.5cm]{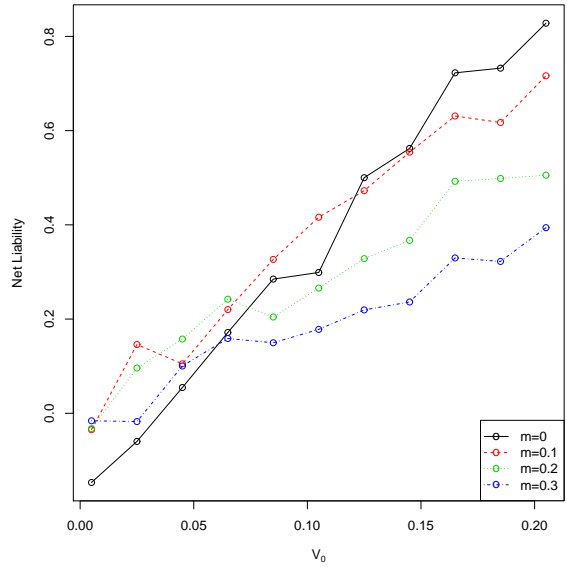}\label{fig:NL_sens_SVJ_wvarying}}
		\\
		\subfigure[Increasing withdrawals]
		{\includegraphics[width=7.55cm]{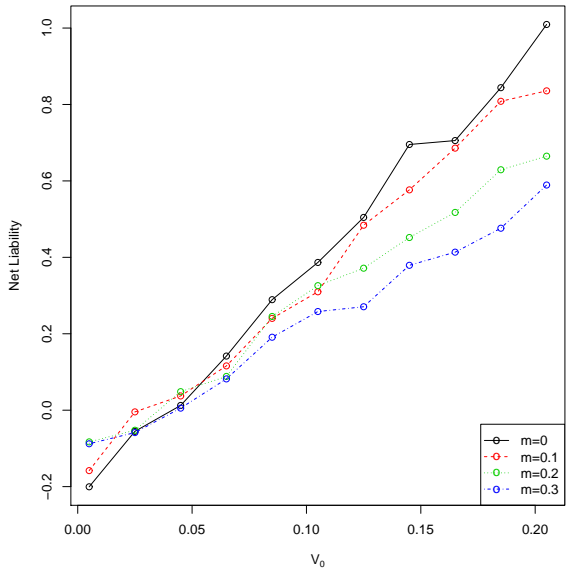}\label{fig:NL_sens_SVJ_wincr}}
		\qquad
		\subfigure[Decreasing withdrawals]
		{\includegraphics[width=7.55cm]{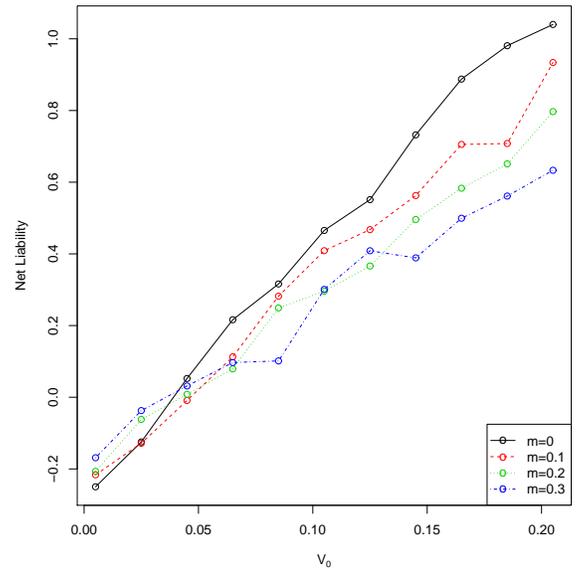}\label{fig:NL_sens_SVJ_wdecr}}	
		\caption{Net liability at $t=0$ as a function of $V_0$.}\label{fig:NL_timevarying}
\end{figure}

Figure \ref{fig:NL_timevarying} presents the net liability of the GMWB contract calculated using the three withdrawal assumptions presented above. In all cases, the fee parameters $\bar{c}$ and $m$ were chosen so that the fee structure is fair, that is, the net liability is null when $V_0 = 0.04$.

Figure \ref{fig:NL_timevarying} shows that the VIX-linked fee reduces the sensitivity of the net liability to $V_0$. That is, the slope of the net liability as a function of $V_0$ decreases significantly as $m$ increases, especially for the deferred withdrawals case. When withdrawals are deferred, an initial high value for $V_0$ leads to higher fee income, while the absence of withdrawals for the first years lets the account value accumulate. The probability that the account is depleted before the guarantee has been withdrawn is thus diminished, compared to the case where the withdrawal start at $t=0$. The impact of the deferred withdrawals is of course more visible when $V_0$ is much larger than $0.04$, the value used to calculate the fair fee. For $V_0 \approx 0.04$, the fair fee structure already takes into account the fact that a certain level of fee income (linked to market voality) is received before the withdrawals start.

Decreasing withdrawals seem to hinder the ability of the VIX-linked fee to reduce the impact of a high value of $V_0$. Indeed, when market are volatile and returns are low, high initial withdrawals will increase the probability that the VA account is emptied before an amount equal to the initial premium is withdrawn. When this is the case, the fee income is also received for a shorter amount of time, and a VIX-linked fee rate will therefore have less impact on the net liability.

\subsection{Net loss}

Our last numerical example focuses on the impact of the VIX-linked fee on the distribution of the discounted future loss (or initial net loss) at inception $\Lambda^W(0,F_0,V_0)$, defined in \eqref{eq:FutureLossW}. The distribution of the initial net loss, especially its tail behavior, is an important part of risk management and risk capital determination (see for example \cite{feng2017risk}). Throughout the section, we use the market parameters given in Table \ref{tab:Parameters} and consider the same contract as in Section \ref{ssec:ExFfee}, unless otherwise stated. For ease of notation, we write $\Lambda^W(0,F_0,V_0) = \Lambda^W_0$.

In this section, we are interested in the real-world ($P$-measure) distribution of the initial loss. For this reason, we simulate the paths of the account value and of the spot volatility using their $P$-measure dynamics described in \eqref{eq:dFt_P}. To do so, we set $\eta_S = 0.6667$, $\eta_V = -2$ and $\eta_J = 1.1414\times 10^{-3}$. Note that we do not differentiate betweeen the jump intensity risk and the jump size risk, and let $\delta^* = \delta$. While the risk-neutral parameters of Table \ref{tab:Parameters} were taken from \cite{kokholm2015joint}, the additional $P$-measure parameter values were chosen arbitrarily, for illustration purposes only. They yield a $P$-measure drift for $S_t$ that is dependent on $V_t$, and equal to $0.0672$ when $V_t = 0.04$. The resulting long-term mean for $V_t$ under the real-world measure is $\frac{\nu}{\varrho^*} = 0.037$. 

Table \ref{tab:stats_loss} presents the $P$-measure expectation, the $P$-measure variance and the CTE at a $90\%$ confidence level for a fixed fee and a VIX-linked fee structure (with $m=0.3$), for various levels of $V_0$. Since the fair fee was calculated using $V_0=0.04$, the results in the table illustrate the effect of a temporary drop ($V_0 = 0.02$) and a temporary increase ($V_0 = 0.08$) in market volatility on the net initial loss $\Lambda^W_0$. For $V_0$ fixed, compared to a fixed fee rate, a VIX-linked fee reduces the expected profit, but leads to a significant drop in the variance of $\Lambda^W_0$. Similarly, even if the mean of the distribution of $\Lambda^W_0$ increases with $m$, its $CTE_{90\%}$ remains relatively stable. Therefore, the VIX-linked fee structure moves the right tail of the initial loss towards the expected initial loss.

Table \ref{tab:stats_loss} also contains the standard deviation of the Monte Carlo estimates (in parentheses below each estimate), calculated using 50 estimates obtained with $N=5\times10^5$ simulations each. The accuracy obtained is very satisfying given that the account value is simulated over a period of 15 years. In such a situation, discretization errors (stemming from the discretization of the integrals) accumulate and affect the variance of the estimate. Further discussion on the accuracy of the Monte Carlo estimates obtained with the explicit and weighted simulation method can be found in \cite{kouritzin2016path}, which also shows that the algorithm compares favorably to other simulation methods.

\renewcommand{\arraystretch}{1.2}
\begin{table}[h!]
	\centering
	
	\caption{Expectation, variance and CTE of the net initial loss $\Lambda_0$ ($m=0.3$ for VIX-linked fee).}
	\label{tab:stats_loss}
	\begin{tabular}{lcccccccc}
		\hline
		~ &~~~& \multicolumn{3}{c}{\textit{Fixed fee}} &~~~& \multicolumn{3}{c}{\textit{VIX-linked fee}} \\
		\hline
		$V_0$ &~~~& ${~~}0.02$ & ${~~}0.04$ & ${~~}0.08$ &~~~& ${~~}0.02$ & ${~~}0.04$ & ${~~}0.08$\\
		\hline
		$E^P[\Lambda^W_0]$ 
		&~~~& $-5.87$ & $-5.95$ & $-6.10$ &~~~& $-4.57$ & $-4.65$ & $-4.84$ \\
		&~~~& \small{$(0.0329)$} & \small{$(0.0313)$} & \small{$(0.0334)$} &~~~& \small{$(0.0288)$} & \small{$(0.0304)$} & \small{$(0.0297)$} \\
		$Var^P[\Lambda^W_0]$ &~~~& $751.10$ & $768.01$ & $801.90$ &~~~& $683.79$ & $698.32$ & $727.65$ \\
		&~~~& \small{$(1.2803)$} & \small{$(1.1935)$} & \small{$(1.3072)$} &~~~& \small{$(1.0334)$} & \small{$(1.0259)$} & \small{$(1.3072)$} \\
		$CTE_{90\%}(\Lambda^W_0)$ &~~~& $39.53$ & $39.91$& $40.70$ &~~~& $39.57$ & $39.90$ & $40.55$\\ 
		&~~~& \small{$(0.0517)$} & \small{$(0.0493)$} & \small{$(0.0526)$} &~~~& \small{$(0.0420)$} & \small{$(0.0406)$} & \small{$(0.0516)$} \\
		\hline
	\end{tabular}
\end{table}

\cite{feng2017risk} also calculate tail-risk measures on the distribution of the loss of a GMWB. They consider fixed fee rates in a market driven by a geometric Brownian motion, which is not calibrated to market data. Comparisons between their results and ours are therefore harder. Nonetheless, we observe that the $CTE_{90\%}$ they obtain for a contract similar to ours (albeit with a lower rider charge and a higher management fee) is in a similar range. In fact, they calculate a $CTE_{90\%}$ of roughly $30\%$ of the initial premium, while ours is closer to 40\%. This could be explained by the presence of stochastic volatility and jumps in our model. These features are known to significantly thicken the left tail of the distribution of the returns, which in turn lead to an increased probability of high losses.

The results of Table \ref{tab:stats_loss} depend on the $P$-measure parameters we chose. To check the sensitivity of our results to the volatility risk premium, we now set $\eta_V = -0.5$ and recalculate the $P$-measure expectation and variance, and the CTE at a $90\%$ confidence level for a fixed fee and a VIX-linked fee structure (with $m=0.3$), for various levels of $V_0$. The results, presented with their standard deviation in Table \ref{tab:stats_loss_sens}, give an indication of the impact of the VIX-linked fee structure on the initial net loss when the absolute value of the volatility risk premium is lower. In this case again, the VIX-linked fee leads to a slightly higher expected loss, with a lower variance.  However, its efficiency in reducing the variance of the loss is diminisehd compared to the case where $\eta_V = -2$.

\begin{table}[h!]
	\centering
	\caption{Expectation, variance and CTE of the net initial loss $\Lambda_0$, ${\eta_V = -0.5}$ ($m=0.3$ for VIX-linked fee).}
	\label{tab:stats_loss_sens}
	
	\begin{tabular}{lcccccccc}
		\hline
		~ &~~~& \multicolumn{3}{c}{\textit{Fixed fee}} &~~~& \multicolumn{3}{c}{\textit{VIX-linked fee}} \\
		\hline
		$V_0$ &~~~& $0.02$ & $0.04$ & $0.08$ &~~~& $0.02$ & $0.04$ & $0.08$\\
		\hline
		$E^P[\Lambda^W_0]$ &~~~& $-7.61$ & $-7.72$ & $-7.97$ &~~~& $-7.07$ & $-7.22$ & $-7.54$ \\
		&~~~& \small{$(0.03778)$} & \small{$(0.0325)$} & \small{$(0.0426)$} &~~~& \small{$(0.0397)$} & \small{$(0.0337)$} & \small{$(0.042)$} \\
		$Var^P[\Lambda^W_0]$ &~~~& $935.40$ & $959.67$ & $1008.11$ &~~~& $898.51$ & $920.39$ & $965.25$ \\
		&~~~& \small{$(1.4652)$} & \small{$(1.7710)$} & \small{$(1.7123)$} &~~~& \small{$(1.2111)$} & \small{$(1.4837)$} & \small{$(1.7877)$} \\
		$CTE_{90\%}(\Lambda^W_0)$ &~~~& $42.59$ & $43.08$& $44.00$ &~~~& $42.20$ & $42.57$ & $43.36$\\ 
		&~~~& \small{$(0.0535)$} & \small{$(0.0501)$} & \small{$(0.0626)$} &~~~& \small{$(0.0488)$} & \small{$(0.0449)$} & \small{$(0.0569)$} \\
		\hline
	\end{tabular}
\end{table}

\section{Conclusion}\label{sec:5_conclusion}

In this paper, we extended the analysis of the VIX-linked fee structure proposed in \cite{cui2016VIX} for VA with GMAB riders to a contract with a GMWB. Our market model includes stochastic volatility and jumps in the index value. We calculated the net liability and the net initial loss associated with the contract using Monte Carlo simulations. Our simulation method is based on an explicit weak solution for the value of the VA account. 

We showed that linking the VA fee rate to the fee structure can help reduce the sensitivity of the net liability to temporary changes in market volatility. We expect that such a fee structure can improve the efficiency of hedging strategies by reducing the variations in the quantity to hedge, i.e. the net liability. We also showed that a VIX-linked fee structure can reduce the impact of a temporary change in market volatility on the tail of the distribution of the initial net loss. This can translate into capital requirements that are less sensitive to spot market volatility.

Throughout the paper, we only consider (possibly time-varying) deterministic policyholder withdrawal behavior. Future research could also consider the impact of a VIX-linked fee structure on lapse and withdrawal incentives.

Another source of future work will be to expand the class of models for which explicit solutions exist, first using the existing techniques and then expanding the techniques to more general (and slightly less efficient) representations.  These expanded models may include double stochastic integrals.

\bibliographystyle{apalike}

\bibliography{References}

\appendix
\setcounter{table}{0}
\renewcommand{\thetable}{A\arabic{table}}

\section{More details on the Proof of Theorem \ref{Theorem2}}

The proof of Theorem \ref{Theorem2} was basically a check that
the given solution was correct.
However, the explicit weak Heston solution introduced in \cite{kouritzin2016path}
was discovered in a more systematic way and new explicit solutions are more
likely to be found by the method in \cite{kouritzin2016path} than by guessing
and checking.
Hence, it is important to explain how the more general solutions
used herein relate to the development in \cite{kouritzin2016path}, which
is the focus of this appendix.
For simplicity, we restrict ourselves to the case $n=2$.
First, it follows by It\^{o}'s formula and the independence of $\widehat \beta,B$
that $ H_t=H^c_t H^i_t$ and $ V_t$ satisfy 
\begin{equation}\label{Heston}
d\!\left(\!\begin{array}{c} H_{t}\\ V_{t}\end{array}\!\right)
=
\left(\!\begin{array}{c}\mu_\kappa H_{t}-\alpha V_t H_{t}\\\nu_\kappa-\varrho  V_{t}\end{array}\!\right)dt+\left(\!\begin{array}{cc}\sqrt{1-\rho^2} H_{t} V_{t}^{\frac{1}{2}}&\rho  H_{t} V_{t}^{\frac{1}{2}}\\0&\kappa  V_{t}^{\frac{1}{2}}\end{array}\!\right)\left(\!\begin{array}{c}dB_{t}\\d \widehat\beta_{t}\end{array}\!\right),
\end{equation}
where we split $H_t^i$, the part of $H_t$ that is independent of $V$, off
\begin{eqnarray}
d\left(\begin{array}{c}H^c_{t}\\V_{t}\end{array}\right)\label{CorrHeston}
&=&\left(\begin{array}{c}(\mu-\alpha V_t)  H^c_{t}\\\nu -\varrho  V_{t}\end{array}\right)dt+\left(\begin{array}{c}\rho H^c_{t} V_{t}^{\frac{1}{2}}\\\kappa  V_{t}^{\frac{1}{2}}\end{array}\right)d\widehat\beta_{t},\\\label{IndHeston}
H^i_t&=&\exp\left(\sqrt{1-\rho^2}\int_0^t  V_s^\frac12 dB_s-\frac{1-\rho^2}2\int_0^t  V_sds\right).
\end{eqnarray}

Now, recall $W^1,W^2$ are independent standard Brownian motions,
set
\begin{equation}\label{ExtendHesSigma}
\sigma(y,z,\tau)=\left[\begin{array}{cc}\frac{\kappa}{2}&0\\0&\frac{\kappa}{2}\\
\rho\,\tau y& \rho\,\tau  z
\end{array}\right]
\end{equation}
and define a new sde of the form:
\begin{equation}\label{ExtendHesa}
d\!\left[\begin{array}{c}Y_t\\Z_t\\H^c_{t}\end{array}\right]
=\left[\begin{array}{c}-\frac{\varrho}2 Y_t\\-\frac{\varrho}2 Z_t\\
(\mu_\kappa-\alpha(Y_t^2+Z^2_t))  H^c_t
\end{array}\right]dt+\!\sigma(Y_t,Z_t,H^c_t)\left[\begin{array}{c}dW^1_{t}\\dW^2_{t}\end{array}\right].
\end{equation}
($Y,Z$ play the role of $Y^1,Y^2$ in the statement of Theorem \ref{Theorem2}.)
This solution can be rewritten as:
\begin{equation}\label{ExtendHes}
d\!\left[\begin{array}{c}Y_t\\Z_t\\H^c_{t}\end{array}\right]
=\left[\begin{array}{c}-\frac{\varrho\,Y_t}2 \\-\frac{\varrho\,Z_t}2 \\(\mu_\kappa-\alpha V_t)  H^c_t\end{array}\right]dt+\!\left[\begin{array}{cc}
\frac{\frac{\kappa}{2}\ Z_t}{\sqrt{Y_t^2+Z_t^2}}&\frac{\frac{\kappa}{2}\ Y_t}{\sqrt{Y_t^2+Z_t^2}}\\
\frac{-\frac{\kappa}{2}\ Y_t}{\sqrt{Y_t^2+Z_t^2}}&\frac{\frac{\kappa}{2}\ Z_t}{\sqrt{Y_t^2+Z_t^2}}\\
0 &\rho H^c_{t} V_{t}^{\frac{1}{2}}\end{array}\right]
\left[\begin{array}{c}dW_{t}\\d\widehat \beta_{t}\end{array}\right],
\end{equation}
where 
\begin{equation}\label{newBMs}
\left[\begin{array}{c}dW_{t}\\d\widehat \beta_{t}\end{array}\right]
=\left[\begin{array}{cc}\frac{Z_t}{\sqrt{Y_t^2+Z_t^2}}&\frac{-Y_t}{\sqrt{Y_t^2+Z_t^2}}\\\frac{Y_t}{\sqrt{Y_t^2+Z_t^2}}&\frac{Z_t}{\sqrt{Y_t^2+Z_t^2}}\end{array}\right]
\left[\begin{array}{c}dW^1_{t}\\dW^2_{t}\end{array}\right].
\end{equation}
Now, \eqref{ExtendHesa} can also be written in terms of Fisk-Stratonovich integration:
\begin{eqnarray}\label{ExtendHesStrat}
\!\!\!\!\!\!\!\!\!d\!\left[\!\begin{array}{c}Y_t\\Z_t\\H^c_{t}\end{array}\!\right]
&\!\!\!=&\!\!\!\!h(Y_t,Z_t,H^c_t)dt
+\!\sigma(Y_t,Z_t,H^c_t)\bullet\!\left[\!\begin{array}{c}dW^1_{t}\\dW^2_{t}\end{array}\!\right]\!,
\end{eqnarray}
where $h$ is the Fisk-Stratonovich drift:
\begin{eqnarray}\label{ExtendStrDrift}
h(y,z,\tau)=\left[\begin{array}{c}-\frac{\varrho}2 y\\-\frac{\varrho}2 z\\\mu  \tau-\frac{\kappa\rho \tau}2-\tau(\frac{\rho^2}2+\alpha)(y^2+z^2)\end{array}\right].
\end{eqnarray}
Now, it is easy to verify that the columns of $\sigma$ of \eqref{ExtendHesSigma} satisfy
$\nabla \sigma_i\sigma_j=\nabla \sigma_j\sigma_i$ so Theorem 2 of 
\citet{Kouritzin/Remillard:2015} can be used to construct a
$\mathcal C^2$-diffeomorphism $\Lambda$ on $\mathbb R\times \mathbb R\times (0,\infty)$ 
that relates local solutions of \eqref{ExtendHesStrat} to local solutions
$\left(\overline Y_t,\overline Z_t,\widehat H^c_t \right)'$ of
the simpler equation
\begin{equation}\label{newh}
d\left[\!\begin{array}{c}\overline Y_t\\\overline Z_t\\\widehat H^c_t\end{array}\!\right]
\! =\! \widehat h(\overline Y_t,\overline Z_t,\widehat H^c_t) dt
+\left[\!\begin{array}{cc}1&0\\0&1\\0&0\end{array}\!\right]d\left[\!\begin{array}{c}W^1_{t}\\W^2_{t}\end{array}\!\right]
\ \text{with } \widehat h =\nabla \Lambda h\circ\Lambda^{-1}	
\end{equation}
via $\left(Y_t,Z_t,H^c_t \right)'=\Lambda^{-1}\left(\overline Y_t,\overline Z_t,\widehat H^c_t \right)'$.
Moreover, the diffeomorphism can be chosen as
$\Lambda =\Lambda_2\circ\Lambda_1$ with $\nabla \Lambda_1\sigma_1\circ\Lambda_1^{-1}=e_1$, 
and $\nabla (\Lambda_2\circ\Lambda_1)\sigma_1\circ(\Lambda_1^{-1}\circ\Lambda_2^{-1})=e_2$,
implying it only depends upon $\sigma$.
Consequently, $\Lambda$ remains the same as in 
\cite{kouritzin2016path}:
\begin{equation}\label{fuldiffeo}
\Lambda(x)=\left[\!\begin{array}{c}\frac2{\kappa}\,x_1\\\frac2{\kappa}\,x_2\\
x_3\exp\left(-\frac\rho\kappa (x_1^2+x^2_2)\right)
\end{array}\!\right],\
\Lambda^{-1}(x)=\left[\!\begin{array}{c}\frac{\kappa}2\,x_1\\\frac{\kappa}2\,x_2\\x_3\exp\left(\rho\frac{\kappa}4
(x_1^2+x^2_2)\right)
\end{array}\!\right],
\end{equation}
and 
\begin{equation}\label{GradLambda}
\!\!\nabla\Lambda (y)=
\!\left[\!\begin{array}{ccc}
\frac2{\kappa}&0&0\\
0&\frac2{\kappa}&0\\
\frac{-2\frac\rho{\kappa}\,y_1y_3}{\exp\left(\frac\rho\kappa (y_1^2+y^2_2)\right)}&
\frac{-2\frac\rho{\kappa}\,y_2y_3}{\exp\left(\frac\rho\kappa (y_1^2+y^2_2)\right)}&
\frac1{\exp\left(\frac\rho\kappa (y_1^2+y^2_2)\right)} \end{array}\right]
\end{equation}
so $\widehat h(x)\doteq (\nabla \Lambda)h\circ\Lambda^{-1}(x)$ satisfies
\begin{eqnarray}\label{hhat}
\widehat h(x)=\!\left[\!\begin{array}{c}
-\frac{\varrho}2 x_1\\
-\frac{\varrho}2 x_2\\
x_3\!\left[\mu -\frac{\kappa\rho}2+\left[\frac{\kappa\rho\varrho}4-\frac{\kappa^2(\rho^2+2\alpha)}8\right]\left\{x_1^2+x_2^2\right\}
\right]
\end{array}\!\!\right].
\end{eqnarray}
The solution for $\left(\overline Y_t,\overline Z_t,\widehat H^c_t \right)'$ 
in \eqref{newh} is:
$\overline Y_t=\int_0^t e^{-\frac{\varrho}2 (t-u)}dW^1_u+e^{-\frac{\varrho}2 t}\overline Y_0$,
$\overline Z_t=\int_0^t e^{-\frac{\varrho}2 (t-u)}dW^2_u+e^{-\frac{\varrho}2 t}\overline Z_0$ (with $\overline Y_0^2+\overline Z_0^2=\frac{\kappa^2}4  V_0$),
and
\begin{equation}\label{Stildeclosed}
\widehat H^c_t=\widehat H^c_0
\exp\bigg(
\left[\mu -\frac{\kappa\rho}2\right] t
+\left[\frac{\kappa\rho\varrho}4-\frac{\kappa^2(\rho^2+2\alpha)}8\right]
\int_0^t
\left\{\overline Y_s^2+\overline Z_s^2\right\} ds
\bigg).
\end{equation}
Moreover, it follows by \eqref{fuldiffeo} and
$\left(Y_t,Z_t,H^c_t \right)'=\Lambda^{-1}\left(\overline Y_t,\overline Z_t,\widehat H^c_t \right)'$ that
\begin{equation}\label{NHb}
H^c_t=\widehat H^c_t
\exp\left(\frac{\rho\kappa}4 (\overline Y_t^2+\overline Z_t^2)\!\right)
=\widehat H_t^c\exp\left(\frac\rho\kappa (Y_t^2+Z_t^2)\!\right)
=\widehat H_t^c\exp\left(\frac\rho\kappa  V_t\!\right)
\end{equation}
and it follows by \eqref{Stildeclosed}, \eqref{NHb}, \eqref{fuldiffeo} and substitution
that
\begin{eqnarray}\label{Snottildeclosed}
\!\!\!\!\!\!\frac{H^c_t}{H^c_0}&\!\!\!=&\!\!\! 
\exp\!\bigg(
\!\left[\mu-\frac{\kappa\rho}2\right] t
+\left[\frac{\kappa\rho\varrho}4-\frac{\kappa^2(\rho^2+2\alpha)}8\right]
\int_0^t\!
\left\{\overline Y_s^2+\overline Z_s^2\right\} ds+\frac\rho\kappa ( V_t- V_0)\!
\bigg)\\\nonumber
\!\!&\!\!\!=&\!\!\!
\exp\!\bigg(
\!\left[\mu-\frac{\kappa\rho}2\right] t
+\left[\frac{\rho\varrho}\kappa-\frac{\rho^2}2-\alpha\right]
\int_0^t V_s ds+\frac\rho\kappa ( V_t- V_0)\!
\bigg).
\end{eqnarray}
Then multiplying $ H_t=H^c_tH^i_t$ with $H^i$ from \eqref{IndHeston}
we get the desired form. \qed

\section{Simulation algorithms}\label{app:simulation_algorithms}

In this section, we present the detailed simulation algorithms discussed in Section \ref{HestSimAlg}. First, it will be notationally convenient to define two constants
\begin{equation*}
	\sigma = \sigma_h=\kappa\sqrt{\frac{1-e^{-\varrho h}}{4\varrho}}\qquad\text{ and }\qquad \psi = \psi_h=e^{-\frac{\varrho h }{2}},
\end{equation*}	
that will be used in the algorithms.

Note that all algorithms take as inputs the parameters pertaining to the market and to the VA contract. We omit to include them as inputs to the algorithm to simplify the presentation.

Subroutine \ref{algo:WeightedOneStep} generates $(\hat{H}_{t_{k}},\hat{V}_{t_{k}},\hat{L}_{t_k})$ from $(H_{t_{k-1}},V_{t_{k-1}},L_{t_{k-1}})$ using the weighted simulation method derived in Section \ref{sec:3_simulation}. Note that $\{\eta^j_\varepsilon\}_{j=1}^{N_{0}}$ will all have been set to $T$ prior to first use, as in Algorithm \ref{algo:SimulationContinuousPaths} (to follow).

Note that the random variables $Z^i$ and $Z^j$ in lines \ref{line:Zi} and \ref{line:Zj} of Subroutine \ref{algo:WeightedOneStep} do not depend upon the inputs like $H$, $L$ and $Y$, so they can be computed a priori offline and just be looked up in real time. When this is done there is a dramatic speed increase.

\begin{subroutinea}[h!]
	\caption{Weighted simulation - One step}\label{algo:WeightedOneStep}
	\begin{algorithmic}[1]
		\Procedure{WeightedSimulation}{$\{H^j_{t_{k-1}},L^j_{t_{k-1}},\eta^j_\varepsilon\}_{j=1}^{N_{t_{k-1}}}$, $\{Y^{j,i}_{t_{k-1}}\}_{j,i=1}^{N_{t_{k-1}},n}$, $h$, $N_{t_{k-1}}$}
		\For{$j$}{1}{$N_{t_{k-1}}$}
		
		\State $\hat{V}^j_{t_k} = 0$
		
		\For{$i$}{1}{$n$}
		\State Generate $(0,1)$-Normal $Z^i$ \label{line:Zi}
		\State $\hat{Y}^{j,i}_{t_k}=\psi Y^{j,i}_{t_{k-1}} + \sigma Z^i$
		\State $\hat{V}^j_{t_k} = \hat{V}^j_{t_k} + (\hat{Y}^{j,i}_{t_{k-1}})^2$ 
		\EndFor
		
		\State $IntV^j = \frac h2 \(V^j_{t_{k-1}} + \hat{V}^j_{t_k}\)$
		\State Generate a $(0,1)$-Normal $Z^j$ \label{line:Zj}
		\State $\hat{H}^j_{t_k} = H^j_{t_{k-1}}
		\exp\left\{a\sqrt{(IntV^j)}Z^j + bh + cIntV^j +d\(\hat{V}^j_{t_k}-V^j_{t_{k-1}}\)\right\}$
		\If {$t_k \leq \eta^j_{\varepsilon}}$
		
		\If{$(V^j_{t_{k-1}} \wedge \hat{V}^j_{t_k}) > \varepsilon$}
		\State$\hat{L}^j_{t_k} = L^j_{t_{k-1}}
		\exp\left\{e\(\log\(\frac{\hat{V}^j_{t_k}}{V^j_{t_{k-1}}}\)+ \varrho h\)+\frac{fh}{2}\(\frac{1}{V^j_{t_{k-1}}} - \frac1{\hat{V}^j_{t_k}}\)\right\}$
		\Else
		\State $\hat{\eta}^j_\varepsilon = t_{k-1}$
		\EndIf
		\EndIf	
		\EndFor
		\EndProcedure
	\end{algorithmic}
\end{subroutinea}

Subroutine \ref{algo:BranchingOneStep} would normally be used in conjunction with Subroutine \ref{algo:WeightedOneStep} when Condition (C) is not satisfied, in order to increase the precision of the results. In this algorithm, $I^j_{t_{k-1}}$ is a vector that tracks the ancestors of the $j^{\text{th}}$ simulated particle. It is updated in lines \ref{line:updateI1} and \ref{line:updateI2} by adding an element to the vector. Note that the algorithm is simplified for ease of presentation; the variables $\eta_{\varepsilon}$ and $\{Y_{t_k}^i\}_{i=1}^n$ also need to be carried along with $\(H_{t_k},V_{t_k},L_{t_k}\)$.

In Subroutine \ref{algo:BranchingOneStep}, $q_1, q_2 > 0$ are parameters that control the amount of resampling, and should be adjusted according to the other parameters used. For more details on the algorithm, the reader is referred to \cite{kouritzin2017branching}, as well as \cite{kouritzin2017convergence,kouritzin2017residual} for more background on these particular branching methods.

\begin{subroutinea}[h!]
	\caption{Branching - One step}\label{algo:BranchingOneStep}
	\begin{algorithmic}[1]
		\Procedure{Branching}{$\{(\hat{H}^j_{t_k}, \hat{V}^j_{t_k}, \hat{L}^j_{t_k}, \eta^j_\varepsilon\}_{j=1}^{N_{t_{k-1}}}$, $\{I^j_{t_{k-1}}\}_{j=1}^{N_{t_{k-1}}}$, $N_{t_{k-1}}$, $q_1$, $q_2$}
		
		\State $A_{t_k} = \frac 1{N_{t_{k-1}}} \sum_{j=1}^{N_{t_{k-1}}} \hat{L}^j_{t_k}$
		\State 
		$r_{t_k} = \exp\left\{q_1
		\(\frac1{N_{t_{k-1}}} \sum_{m=1}^{N_{t_{k-1}}} \(\log \hat{L}^m_{t_k}\)^2 
		-\(\frac1{N_{t_{k-1}}} \sum_{m=1}^{N_{t_k}} \log \hat{L}^m_{t_k}\)^2 \)^{\frac{q_2}{2}} \right\}$ 	
		\State Set $l = 0$	
		\For{$m$}{1}{$N_{t_{k-1}}$}
		\If {$\hat{L}^m_{t_k} \notin \left(\frac{A_{t_k}}{r_{t_k}}, A_{t_k}r_{t_k}\right)}$
		\State $(\hat{H}^{m-l}_{t_k},\hat{V}^{m-l}_{t_k},\hat{L}^{m-l}_{t_k}) 
		= (\hat{H}^{m}_{t_k},\hat{V}^{m}_{t_k},\hat{L}^m_{t_k})$
		\State $\hat{I}^{m-l}_{t_k} = I^m_{t_{k-1}}$
		\Else
		\State $l = l+1$
		\State $(H^l_{t_k},V^{l}_{t_k},L^{l}_{t_k})
		= (\hat{H}^{m}_{t_k},\hat{V}^{m}_{t_k},\hat{L}^m_{t_k})$
		\State $I^l_{t_k} = (I^m_{t_{k-1}}, l)$ \label{line:updateI1}
		\EndIf
		\EndFor	
		\State $N_{t_k} = l$
		\For{$m$}{1}{$(N_{t_{k-1}}-l)$}
		\State Generate $\[\frac{m-1}{N_{t_{k-1}}-l},\frac{m}{N_{t_{k-1}}-l}\]$-Uniform $W_{t_k}^{m+l}$
		\EndFor
		\State $\{U^m_{t_k} = W^{p(m)}_{t_k}\}_{m=1}^{N_{t_{k-1}}-l}$, for $p$ a random permutation of $\{l+1,\ldots,N_{t_{k-1}}\}$
		\For{$m$}{$l+1$}{$N_{t_{k-1}}$}
		\State $N^m_{t_k} = \lfloor\frac{\hat{L}^{m-1}_{t_k}}{A_{t_k}}\rfloor
		+ \1_{\left\{U^k_{t_k} \leq \(\frac{\hat{L}^{m-1}_{t_k}}{A_{t_k}} - \lfloor\frac{\hat{L}^{m-1}_{t_k}}{A_{t_k}}\rfloor\)\right\}}$	
		\For{$j$}{1}{$N^m_{t_k}$}
			\State $\(H^{N_{t_k}+j}_{t_k},V^{N_{t_k}+j}_{t_k},L^{N_{t_k}+j}_{t_k}\) = \(\hat{H}_{t_k}^{m-l},\hat{V}_{t_k}^{m-l},A_{t_k}\)$
			\State $I^{N_{t_k}+j}_{t_k} = \(\hat{I}^{m-l}_{t_k},N_{t_k}+1\)$ \label{line:updateI2}
		\EndFor
		\State $N_{t_k} = N_{t_k} + N^m_{t_k}$
		\EndFor
		\EndProcedure
	\end{algorithmic}
\end{subroutinea}

Algorithm \ref{algo:SimulationContinuousPaths} simulates paths of $(H_t,V_t,L_t)_{t=0}^T$ by calling Subroutines \ref{algo:WeightedOneStep} and \ref{algo:BranchingOneStep}.

\begin{algorithm}
	\caption{Simulation of paths of $(H_t,V_t,L_t)_{t=0}^T$}\label{algo:SimulationContinuousPaths}
	\begin{algorithmic}[1]
		\Procedure {SimulationContinuousPaths}{$S_0$, $V_0$, $N_0$, $h$, $T$}
		\State $\{(H^j_0,V^j_0,L^j_0,\eta^j_\varepsilon,I^j_0) = (S_0,V_0,1,T,1)\}_{j=1}^{N_0}$,
		$\{Y_0^{i,j} = \sqrt{(\frac{V_0}{n})}\}_{j,i=1}^{N_0,n}$
		\State $K = \lfloor \frac Th \rfloor$	
		\State $\{I^j_0\}_{j=1}^{N_0} = \emptyset$
		\For{k}{1}{$K$}
		\State Simulate $\{(\hat{H}^j_{t_k},\hat{V}^j_{t_k},\hat{L}^j_{t_k})\}_{j=1}^{N_{t_{k-1}}}$ using Subroutine \ref{algo:WeightedOneStep}.
		\State Branch particles to obtain $\{(H^j_{t_k},V^j_{t_k},L^j_{t_k},\eta^j_\varepsilon,I^j_{t_k})\}_{j=1}^{N_{t_k}}$ using Subroutine \ref{algo:BranchingOneStep}.
		\EndFor
		\EndProcedure
	\end{algorithmic}
\end{algorithm}

After having simulated $\{(H^j_{t_k},V^j_{t_k},L^j_{t_k})\}_{j,k=1}^{N_{t_k},K}$, with $K = \lfloor \frac Th \rfloor $, $\{I^j_{t_K}\}_{j=1}^{N_{t_K}}$, will be a ${(N_{t_K} \times K)}$ matrix tracking the ancestors of the simulated values $\{(H^j_{t_K}, V^j_{t_K})\}_{j=1}^{N_{t_K}}$. Going forward, we write $\{I^j_{t_K}\}_{j=1}^{N_{t_K}} = I_{t_K}$, and we denote its $k^{\text{th}}$ column by $I_{t_K}|_k$. This matrix will be used in Subroutine \ref{algo:jumps} and Algorithm \ref{algo:NetLiability} to keep track of the historical paths.

\begin{subroutineb}
	\caption{Addition of jumps - One step}\label{algo:jumps}
	\begin{algorithmic}[1]
		\Procedure{Jumps}{$H_{t_{k-1}}^{j_1}$, $G_{t_{k-1}}^{j_1}$, $H^{j_2}_{t_k}$}
		
			\State $G^{j_2}_{t_k} =  
				G^{j_1}_{t_{k-1}} H^{j_2}_{t_k}/H^{j_1}_{t_{k-1}}$
			\State Generate a $(\lambda h)$-Poisson $M$
			\For{$i$}{1}{$M$}
					
				\State Generate a $\(\log(1+\delta)-\frac{\chi^2}{2},\chi^2\)$-Normal  $Y^{i}$
				\State $G^{j_2}_{t_k} = G^{j_2}_{t_k} e^{Y^i}$
										
			\EndFor
		
		\EndProcedure
	\end{algorithmic}
\end{subroutineb}

Algorithm \ref{algo:NetLiability} is used to calculate the net liability associated with the GMWB presented in Section \ref{ssec:GMWB}. The algorithm takes as input the paths produced by Algorithm \ref{algo:SimulationContinuousPaths}, as well as a vector $w$ containing the withdrawal rates for each period. It is possible to use more precise integral approximations in lines \ref{line:approxR} and \ref{line:approxC} to improve the performance of the algorithm. With minor modifications, Algorithm \ref{algo:NetLiability} can also be used to calculate the risk measures given in \ref{ssec:riskmeasures}.

\begin{algorithm}
\caption{Net liability calculation}\label{algo:NetLiability}
\begin{algorithmic}[1]
	\Procedure{GMWB}{$\{(G_{t_k}^j,V_{t_k}^j)\}_{j,k=1,0}^{N_{t_k},K}$, $\{(L^j_{t_K},I^j_{t_K})\}_{j=1}^{N_{t_K}}$}
	
	\State $\{(G^j_0,F^j_0,R^j_0,C^j_0,W^j_0)=(1,F_0,0,0,0)\}_{j=1}^{N_0}$	
	\State $K = \lfloor \frac{T}{h} \rfloor$
	\State $NetLiability = 0$
	
	\For{$k$}{1}{$K$}	
		
		\State $D = \emptyset$
		\For{$l$}{1}{$N_{t_K}$}	
		\If{$I^l_{t_K}|_k \notin D$}				
			\State $j = I^l_{t_K}|_{k-1}$, $m = I^l_{t_K}|_k$ 
				
			\If{$F^{j}_{t_{k-1}} > 0$}				
					
				\State Generate $G^{m}_{t_k}$ using Subroutine \ref{algo:jumps}
				\State $R^{m}_{t_k} =  
					R^{j}_{t_{k-1}} + 
					\frac h2 \(\frac{w_{t_{k-1}}}{G^{j}_{t_{k-1}}}
					+\frac{w_{t_k}}{G^{m}_{t_k}}\)$
					\label{line:approxR}
				\State $F^{m}_{t_k} = \max\(G^{m}_{t_k}[1-R^{m}_{t_k}],0\)$	
				\State $C^{m}_{t_k} =	 C^{j}_{t_{k-1}} + 	\frac{h}{2}e^{-rht_k} 	
					\label{line:approxC}
					[(\alpha_0 - q + \alpha V^{j}_{t_{k-1}})F^{j}_{t_{k-1}} e^{rh}
					+(\alpha_0 - q + \alpha V^{m}_{t_k})F^{m}_{t_k}]$
				\State $W^{m}_{t_k} = 0$
				
			\Else
					
				\State $C^{m}_{t_k} = C^{j}_{t_{k-1}}$
				\State $W^{m}_{t_k} = W^{j}_{t_{k-1}} 
					+ \frac{he^{-rht_k}}{2}\[w_{t_{k-1}}e^{rh} + w_{t_k}\]$
				
			\EndIf
				
			\State $D = D \cup I^l_{t_K}|_k$
			
			\If{$k=K$}
			
				\State $NetLiability = \frac{l-1}{l}NetLiability + \frac1l (W^m_{t_K}-C^m_{t_K}) L^m_{t_K}$
			
			\EndIf

		\EndIf			
		\EndFor
	\EndFor

	\EndProcedure
\end{algorithmic}
\end{algorithm}

\end{document}